\documentclass[11pt,a4paper]{article}
\usepackage{jinstpub}
\usepackage{newtxtext}
\usepackage{amsthm} 
\usepackage{newtxmath}
\usepackage[zerostyle=e]{newtxtt}
\usepackage{amsmath}
\usepackage{amssymb}
\usepackage{graphicx}
\usepackage{wrapfig}

\usepackage[T1]{fontenc}
\usepackage{lineno}
\usepackage{color}
\usepackage{svn-multi}
\usepackage{mathtools}
\usepackage{placeins}
\usepackage{rotating}
\notoc

\begin{document}


\title{Performance of the EDELWEISS-III experiment for direct dark matter searches}

\collaboration{EDELWEISS collaboration}

\author[a]{E.~Armengaud,}
\author[b,1]{Q.~Arnaud,%
\note{Now at Physics Department, Queen's University, Kingston, ON, Canada.}}
\author[b]{C.~Augier,}
\author[c]{A.~Beno\^{i}t,}
\author[d]{L.~Berg\'{e},}
\author[e]{T.~Bergmann,}
\author[b]{J.~Billard,}
\author[a]{T.~de~Boissi\`{e}re,}
\author[c]{G.~Bres,}
\author[d,f]{A.~Broniatowski,}
\author[g]{V.~Brudanin,}
\author[c]{P.~Camus,}
\author[b]{A.~Cazes,}
\author[d]{M.~Chapellier,}
\author[b]{F.~Charlieux,}
\author[b]{M.~De~J\'{e}sus,}
\author[d]{L.~Dumoulin,}
\author[h]{K.~Eitel,}
\author[g]{D.~Filosofov,}
\author[f]{N.~Foerster,}
\author[a]{N.~Fourches,}
\author[c]{G.~Garde,}
\author[b]{J.~Gascon,}
\author[d]{A.~Giuliani,}
\author[c]{M.~Grollier,}
\author[a]{M.~Gros,}
\author[h,2]{L.~Hehn,%
\note{Now at Nuclear Science Division, Lawrence Berkeley National Laboratory, Berkeley, CA, US.}}
\author[a]{S.~Herv\'{e},}
\author[f]{G.~Heuermann,}
\author[d]{V.~Humbert,}
\author[i]{Y.~Jin,}
\author[b]{A.~Juillard,}
\author[b,f,3]{C.~K\'{e}f\'{e}lian,%
\note{Now at Physics Department, University of California, Berkeley, CA, US.}}
\author[e]{M.~Kleifges,}
\author[f]{V.~Kozlov,}
\author[j]{H.~Kraus,}
\author[k]{V.~A.~Kudryavtsev,}
\author[d]{H.~Le-Sueur,}
\author[j,4]{J.~Lin,}
\author[b]{R.~Maisonobe,}
\author[d]{M.~Mancuso,}
\author[d]{S.~Marnieros,}
\author[e]{A.~Menshikov,}
\author[a]{X.-F.~Navick,}
\author[a]{C.~Nones,}
\author[d]{E.~Olivieri,}
\author[l]{P.~Pari,}
\author[a]{B.~Paul,}
\author[d]{D.~Poda,}
\author[b]{E.~Queguiner,}
\author[k]{M. Robinson,}
\author[c]{H.~Rodenas,}
\author[g]{S.~Rozov,}
\author[b]{V.~Sanglard,}
\author[f,2]{B. Schmidt,}
\author[f,4]{S.~Scorza,%
\note{Now at SNOLAB, Lively, ON, Canada.}}
\author[h]{B.~Siebenborn,}
\author[e]{D.~Tcherniakhovski,}
\author[b]{L.~Vagneron,}
\author[e]{M.~Weber,}
\author[g]{E.~Yakushev,}
\author[j,1]{X.~Zhang,}
\author[a]{A.~Zolotarova}
\affiliation[a]{IRFU, CEA, Universit\'{e} Paris-Saclay, F-91191 Gif-sur-Yvette, France}
\affiliation[b]{Univ Lyon, Universit\'{e} Lyon 1, CNRS/IN2P3, IPN-Lyon, F-69622, Villeurbanne, France}
\affiliation[c]{Institut N\'{e}el, CNRS/UJF, 25 rue des Martyrs, BP 166, 38042 Grenoble, France}
\affiliation[d]{CSNSM, Univ. Paris Sud, CNRS/IN2P3, Universit\'{e} Paris-Saclay, 91405 Orsay, France}
\affiliation[e]{Karlsruher Institut f\"{u}r Technologie, Institut f\"{u}r Prozessdatenverarbeitung und Elektronik, Postfach 3640, 76021 Karlsruhe, Germany}
\affiliation[f]{Karlsruher Institut f\"{u}r Technologie, Institut f\"{u}r Experimentelle Kernphysik, Gaedestr. 1, 76128 Karlsruhe, Germany}
\affiliation[g]{JINR, Laboratory of Nuclear Problems, Joliot-Curie 6, 141980 Dubna, Moscow Region, Russian Federation}
\affiliation[h]{Karlsruher Institut f\"{u}r Technologie, Institut f\"{u}r Kernphysik, Postfach 3640, 76021 Karlsruhe, Germany}
\affiliation[i]{Laboratoire de Photonique et de Nanostructures, CNRS, Route de Nozay, 91460 Marcoussis, France}
\affiliation[j]{University of Oxford, Department of Physics, Keble Road, Oxford OX1 3RH, UK}
\affiliation[k]{University of Sheffield, Department of Physics and Astronomy, Sheffield, S3 7RH, UK}
\affiliation[l]{DSM/IRAMIS, CEA, Universit\'{e} Paris-Saclay, F-91191 Gif-sur-Yvette, France}
\emailAdd{claudia.nones@cea.fr}
\emailAdd{c.augier@ipnl.in2p3.fr}

\abstract{
We present the results of measurements demonstrating the efficiency  of the EDELWEISS-III array
of cryogenic germanium detectors for direct dark matter searches. The experimental setup and the FID (Fully Inter-Digitized) detector array is described, as well as the efficiency of the double measurement of heat and ionization signals in background rejection. For the whole set of 24~FID detectors used for coincidence studies, the baseline resolutions for the fiducial ionization energy are mainly below 0.7~keV$_{ee}$ (FHWM) whereas the baseline resolutions for heat energies are mainly below 1.5~keV$_{ee}$ (FWHM). The response to nuclear recoils as well as the very good discrimination capability of the FID design has been measured with an AmBe source. The surface $\beta$- and $\alpha$-decay rejection power of $R_{\rm surf} < 4 \times 10^{-5}$ per $\alpha$ at 90\%\,C.L. has been determined with a $^{210}$Pb source, the rejection of bulk $\gamma$-ray events has been demonstrated using $\gamma$-calibrations with $^{133}$Ba sources leading to a value of $R_{\gamma{\rm -mis-fid}} < 2.5 \times 10^{-6}$ at 90\% C.L.. The current levels of natural radioactivity measured in the detector array are shown as the rate of single $\gamma$ background. The fiducial volume fraction of the FID detectors has been measured to a weighted average value of $(74.6 \pm 0.4)\%$ using the cosmogenic activation of the $^{65}$Zn and $^{68,71}$Ge isotopes. The stability and uniformity of the detector response is also discussed. The achieved resolutions, thresholds and background levels of the upgraded EDELWEISS-III detectors in their setup are thus well suited to the direct search of WIMP dark matter over a large mass range.
}

\keywords{Dark Matter detectors (WIMPs, axions, etc.), Cryogenic detectors, Solid state detectors, Large detector-systems performance}


\maketitle


\newpage

\svnid{$Id: introduction.tex 114 2017-05-23 14:52:18Z caugier $}
\section{Introduction}
\label{sec:intro}

There is strong evidence in favour of the existence of non-baryonic dark matter at almost every cosmic scale~\cite{DMevidence, DMevidence2}. Theories and simulations regarding hierarchical structure formation indicate that this non-luminous component may manifest itself in the form of a gas of Weakly Interacting Massive Particles (WIMPs)~\cite{dmreview}. There is no viable candidate in the Standard Model of particle physics for the composition of this cold dark matter gas. There are however theories beyond the Standard Model, specifically developed to solve problems inherent to elementary particle physics, that lead to very attractive dark-matter candidates in the form of WIMPs. Natural and popular candidates, like the supersymmetric neutralino, have a predicted mass in the range of a few GeV/c$^2$ to  TeV/c$^2$ and an elastic scattering cross section on nucleons at or below the weak scale~\cite{dmSUSY}. Furthermore, they are characterized by a dominant interaction with atomic nuclei, inducing therefore low-energy nuclear recoils in the target material, and could be thus detected in the so-called direct detection experiments~\cite{dmdirect}. There is an intense experimental activity on dark matter direct detection since many years, for which the most promising results have been obtained with liquid noble or cryogenic detectors: see e.g.~\cite{xenon10-TP, xenon100-TP, LUX-TP, PandaX-TP, DarkSide50, cresstII, cdmsII-2015, DAMA-LIBRA-TP} and references therein for information on their technical designs and performance.

The goal of the EDELWEISS experiment is to perform a direct search for WIMPs from the galactic halo using an array of cryogenic germanium detectors able to identify events consisting of WIMP-induced nuclear recoils with a kinetic energy in the keV to tens of keV range. The main challenge for this rare-event search experiment is to distinguish a potential WIMP signal from recoils induced by natural radioactivity and cosmic rays.

\begin{figure}
\begin{center}
\includegraphics[width=0.8\textwidth]{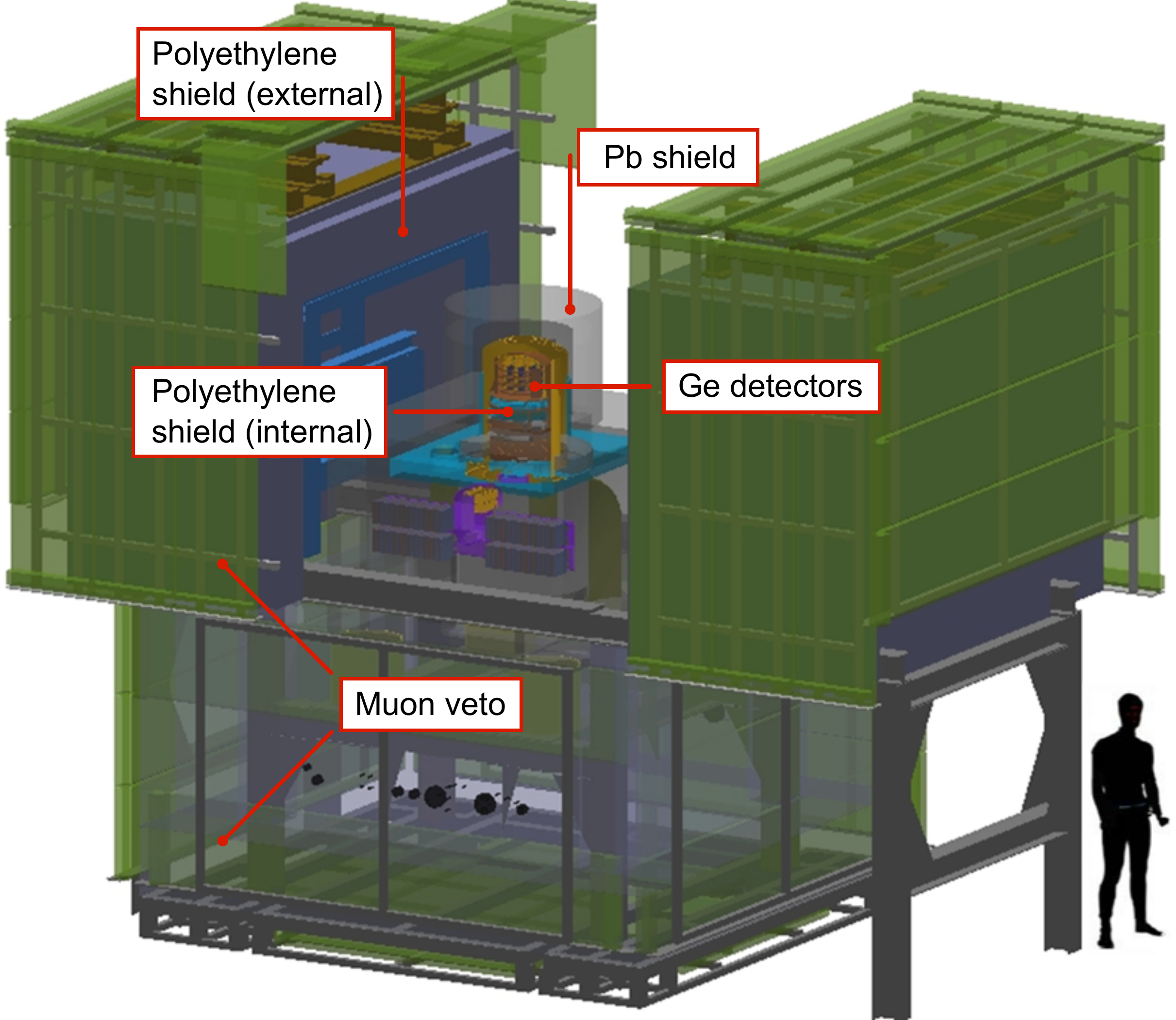}
\caption{Schematic view of the EDELWEISS-III setup showing in the center the cryostat
hosting the germanium bolometers, surrounded by passive lead (Pb) and
polyethylene shields and an active muon veto in order to protect the
detectors from various backgrounds.}
\label{fig:EDW-All-Simu-Text}
\end{center}
\end{figure}

The EDELWEISS detectors, working at 18~mK, are hosted inside a cryostat and operated in a radiopure underground environment located in the Modane Underground Laboratory (Laboratoire Souterrain de Modane, LSM) in France. As shown in figure~\ref{fig:EDW-All-Simu-Text}, shields of polyethylene (PE) and lead, as well as an active muon veto, surround the detector setup allowing both passive and active background rejection. The remaining dominant background, due to $\beta$ and $\gamma$ natural radioactivity, is suppressed by using the capability of the germanium detectors of clearly separating the electron recoils induced by $\beta$ and $\gamma$ radiation from nuclear recoils, which are potential signals, using a double-readout event-by-event discrimination: a calorimetric measurement of the total deposited energy and the ionization yield. At the operating temperature of 18~mK the nuclear recoils induce a temperature increase of roughly 0.1~$\mu$K per 1~keV that is measured with heat sensors. The ionization signal is collected on electrodes covering all the surfaces.

In 2009 and 2010, the EDELWEISS collaboration conducted a WIMP search  using an array of ten
400~g Inter-Digitized (ID) detectors~\cite{edw2}. It was found that the remaining events after rejection were compatible with the ionization yield of electron recoils and revealed the limitations of the ID design. The results of the background studies performed in the context of EDELWEISS-II have highlighted a need of improving the shielding. It was also shown that the primary source of $\gamma$ background in EDELWEISS-II originated from a range of copper elements in the vicinity of the detectors, whereas neutron background was mainly dominated by the $(\alpha,n)$ reactions in electronic components inside the lead shielding~\cite{EDW-background}.

The initial goal of EDELWEISS-III was to probe WIMP-nucleon cross sections down to $10^{-9}$ pb for a WIMP mass of $\approx$\,50~GeV with 12\,000{$\, \rm{kg \cdot d}$} exposure.
Many technical upgrades were made to reduce the residual background to less than one event for one year of data taking. With the aim of increasing the detection efficiency for low-mass WIMPs at [$\approx$\,5,20]~GeV, both the electronics (readout and DAQ) and the cryogenic systems have been modified to improve the energy resolutions and subsequently lower the energy threshold. These changes have been coupled to a new germanium bolometer design, called FID (Fully Inter-Digitized) detector, that allows to reach very good active background rejection as described in this paper.

The detector characterization being the focus of this paper is based on data obtained during a period of 10 months. For this run (Run~308) 36~FID detectors were installed in the EDELWEISS cryostat in a tower array configuration. This long run, for which only 24 detectors were read out, was focused on WIMP search, with frequent calibrations and several tests in between. The studies presented here were performed on a data set which was not blinded for the WIMP search analysis, notably multiple hit events, events outside the fiducial volume and fiducial events with an ionization yield larger than 0.5. 
Shorter dedicated data taking periods have been used to demonstrate the rejection power efficiency or the high-voltage capabilities of the FID design.

The present paper is organized as follows: sections~\ref{sec:detectors} to \ref{sec:shields} focus on the general description of the detectors, cryogenics and shielding, then sections~\ref{sec:readout} and \ref{sec:daq} describe the read out and acquisition parts. It is followed by an overview of the detector performance in terms of resolution and active background rejection (section~\ref{sec:performance}), while the efficiencies of the shielding and muon-veto are given in section~\ref{sec:array-performance}. Concluding remarks and prospects can be found in section~\ref{sec:conclusion}.

\svnid{$Id: detectors.tex 103 2017-05-02 12:47:54Z keitel $}
\section{EDELWEISS FID detectors}
\label{sec:detectors}

Detectors used in the EDELWEISS-III experiment, called FID, are made of ultrapure germanium cylindrical crystals with a height of 4~cm, a diameter of 7~cm and a mass of 820--890~g.\footnote{Crystals are from CANBERRA Industries, see http://www.canberra.com/products/, or Baltic Scientific Instruments (BSI), see http://bsi.lv/en/. No specific cleaning has to be carried out on the crystals produced by these two companies.} These detectors are instrumented to perform a double heat-and-ionization measurement of signals arising from particle interactions. The double read-out is used to discriminate background-induced electron recoils (ER) from potential WIMP-induced nuclear recoils (NR) on an event-by-event basis.

Heat signals are measured with two Neutron Transmutation Doped (NTD) Ge sensors~\cite{ntd} glued on the center of the top and bottom surfaces of the crystal thanks to a veil of Araldit glue. Their working principle and characteristics are given in section~\ref{sec:sub:heat-channel}. NTD of two different productions are used with geometries of $4\times4\times0.45$ mm$^3$ and $3\times5\times1$ mm$^3$ respectively. Charge signals are measured with a set of evaporated electrodes.

\begin{figure}[h]
\begin{center}
\includegraphics[width=0.44\textwidth]{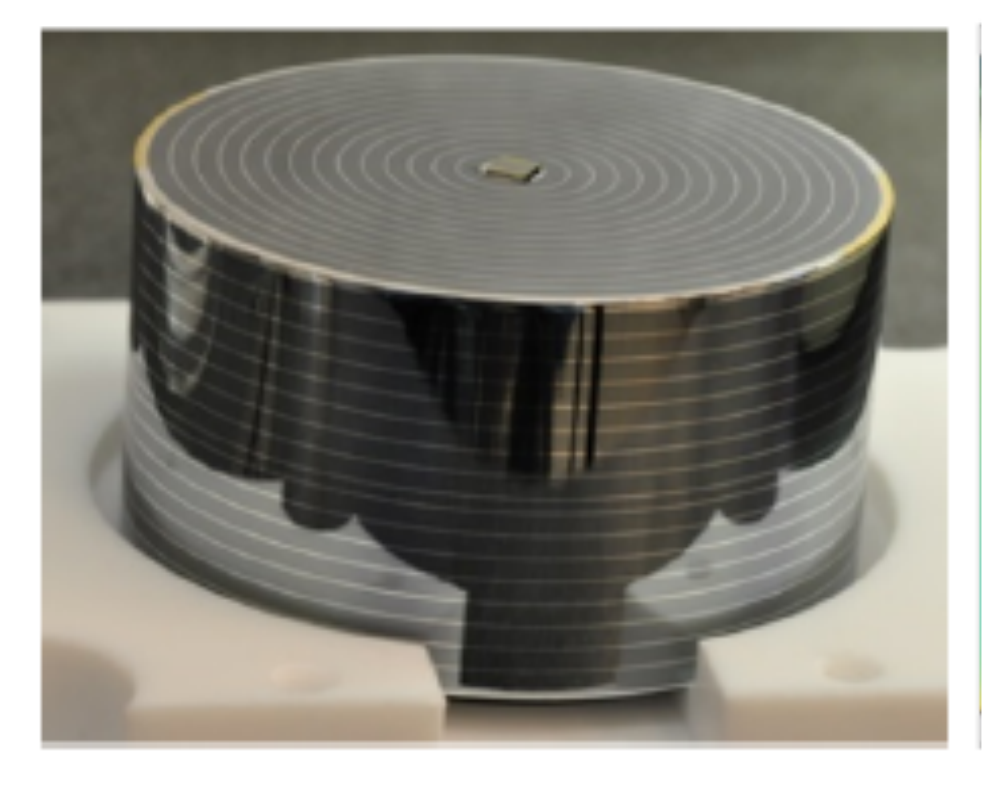}
 \includegraphics[width=0.52\textwidth]{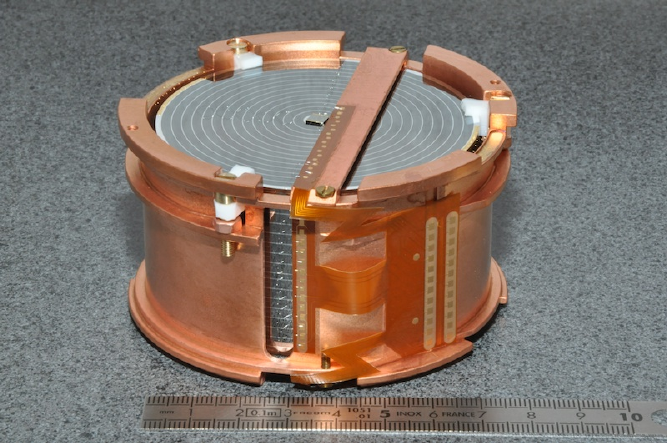}
\caption{Images of an 800-g FID detector with concentric ring electrodes covering the entire crystal surface and one of the two NTD sensors glued onto the top surface. On the right photo, the bolometer is surrounded by its copper casing and held by Teflon clamps.}
\label{fig:FID800-picture}
\end{center}
\end{figure}

The dual heat-ionisation measurement allows to discriminate ER vs NR events when charge collection is complete, as described in section~\ref{sec:performance}. In addition, to reject interactions taking place close to the crystal surfaces, a special concentric-electrode scheme, based on the co-planar grid technique for event localization~\cite{id}, has been first developed and tested in the EDELWEISS-II experiment with an array of ten 400-g ID detectors~\cite{edw2, idfirst}. These are heat-and-ionization cryogenic detectors equipped with interleaved electrodes on flat surfaces for the rejection of near-surface events. The same concentric-electrode scheme has been improved for FID detectors, as shown in figure~\ref{fig:FID800-picture}: 200~nm thick Al electrodes are evaporated on the whole Ge crystal, both flat and side surfaces, in the form of annular concentric rings 150~$\mu$m  wide with a 2~mm pitch. A surface passivation treatment, with a 60 to 80~nm amorphous layer of hydrogenated Ge is deposited only under the electrodes, the surface between electrodes being left unprocessed. To reduce possible residual leakage current, a preventive post-processing XeF$_{2}$ pulsed dry etching of the detector is applied~\cite{XeF}.

Each detector is fully surrounded by a low-radioactivity copper casing to protect it from infrared radiation and is held by Teflon clamps (see figure~\ref{fig:FID800-picture}, right). The thermal connection to the bath needed to return the detectors to equilibrium is formed through a gold pad directly evaporated onto the crystal with typically 20 Au wires of 25~$\mu$m diameter each.

To reduce possible surface contamination, all copper and PTFE pieces are cleaned before the detector installation with the following procedure: cleaning with ethanol, etching with a solution prepared using nitric acid and ultrapure water, rinsing with deionized ultrapure water, drying using industrial dryer. Concerning PTFE pieces, only the cleaning step with ethanol is carried out.


The concentric electrodes are connected electrically such that all odd-numbered rings are connected to each other as well as all even-numbered rings. With two sets of rings on the top half and the other two on the bottom half, biasing requires four voltages to measure the charge. This full coverage of the ultra-pure Ge crystals with interleaved electrodes, connected by ultrasonic wedge bonding with Al wires, is a key feature of the EDELWEISS FID detectors. In contrast to a planar scheme, when voltages are set, interleaved electrodes induce much higher electric fields along both flat and cylindrical surfaces allowing an efficient discrimination between bulk and surface events.

\begin{figure}[ht]
\begin{center}
\includegraphics[width=0.75\textwidth]{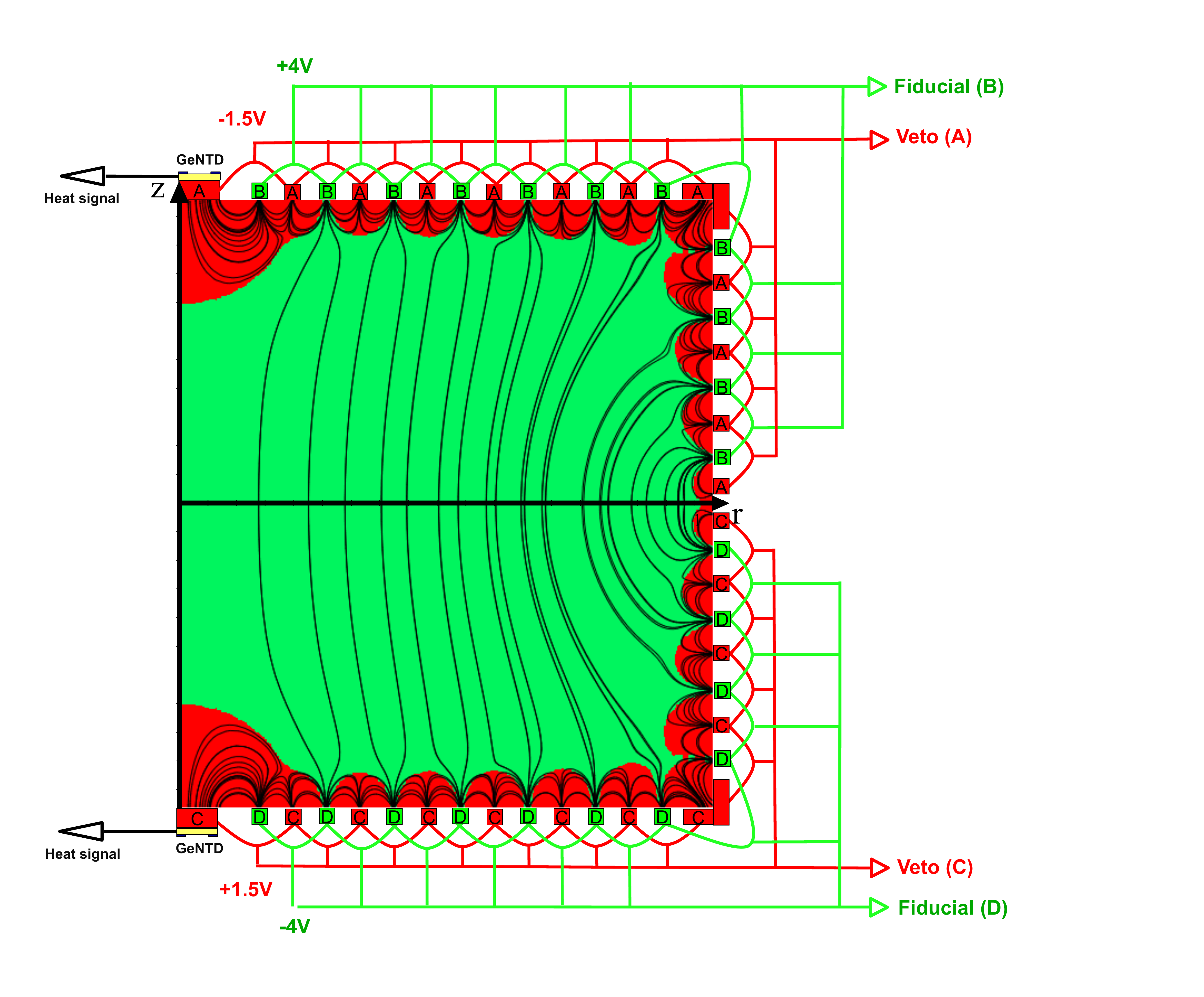}
\caption[]{Cross-section of the FID detector design showing the interleaved electrode scheme. In a standard configuration, $B$ and $D$ collecting electrodes are biased at V$_{B} = + 4$~V and V$_{D} = - 4$~V and define the fiducial volume (green region). $A$ and $C$ veto electrodes are biased at V$_{A} = - 1.5$~V and V$_{C} = + 1.5$~V and define the surface volume (red region). In the bulk of the detector, the field lines are nearly vertical. Near the surfaces they are parallel to the surfaces. }
\label{fig:FID800sheme_1}
\end{center}
\end{figure}

As explained in sections~\ref{sec:sub:Calib} and \ref{sec:fiducialcuts}, by applying appropriate selection cuts on both veto and fiducial channels, interactions which take place in the regions of detectors highlighted in red in figure~\ref{fig:FID800sheme_1} can be identified. Events occurring near the border between the surface and bulk region exhibit charge sharing between more than two measurement channels and are also tagged and eventually rejected by these selection rules.

Typical voltage settings for FID detectors are $\pm 4$~V and $\mp1.5$~V. This corresponds to a range high enough to provide sufficient electric field inside the crystal for charge drift~\cite{charge-paper}, but low enough in order not to spoil the ER rejection due to the Neganov-Luke effect~\cite{neganov, luke} (see section~\ref{sec:Ionization-heat-recoil-energies}). The limit imposed by leakage current on the highest applicable voltage varies from detector to detector. Two of the detectors (see Table~\ref{tab:baseline-resol-threshold}) are limited to  $\pm 3.2$~V while some others could be operated at $\pm 50$~V in Neganov-Luke-boost mode (see section~\ref{subsec:luke-neganov}). A specific surface treatment has been developed~\cite{XeF} in order to ensure that all but two detectors have smaller than 1~fA current at the standard voltage $\pm 4$~V and that the typical voltage limit across any pair of adjacent electrodes is greater than $15$~V.

\svnid{$Id: cryogenics.tex 114 2017-05-23 14:52:18Z caugier $}

\section{Cryogenic system}
\label{sec:cryogenics}

\begin{figure}[h]
\begin{center}
\includegraphics[width=1\textwidth]{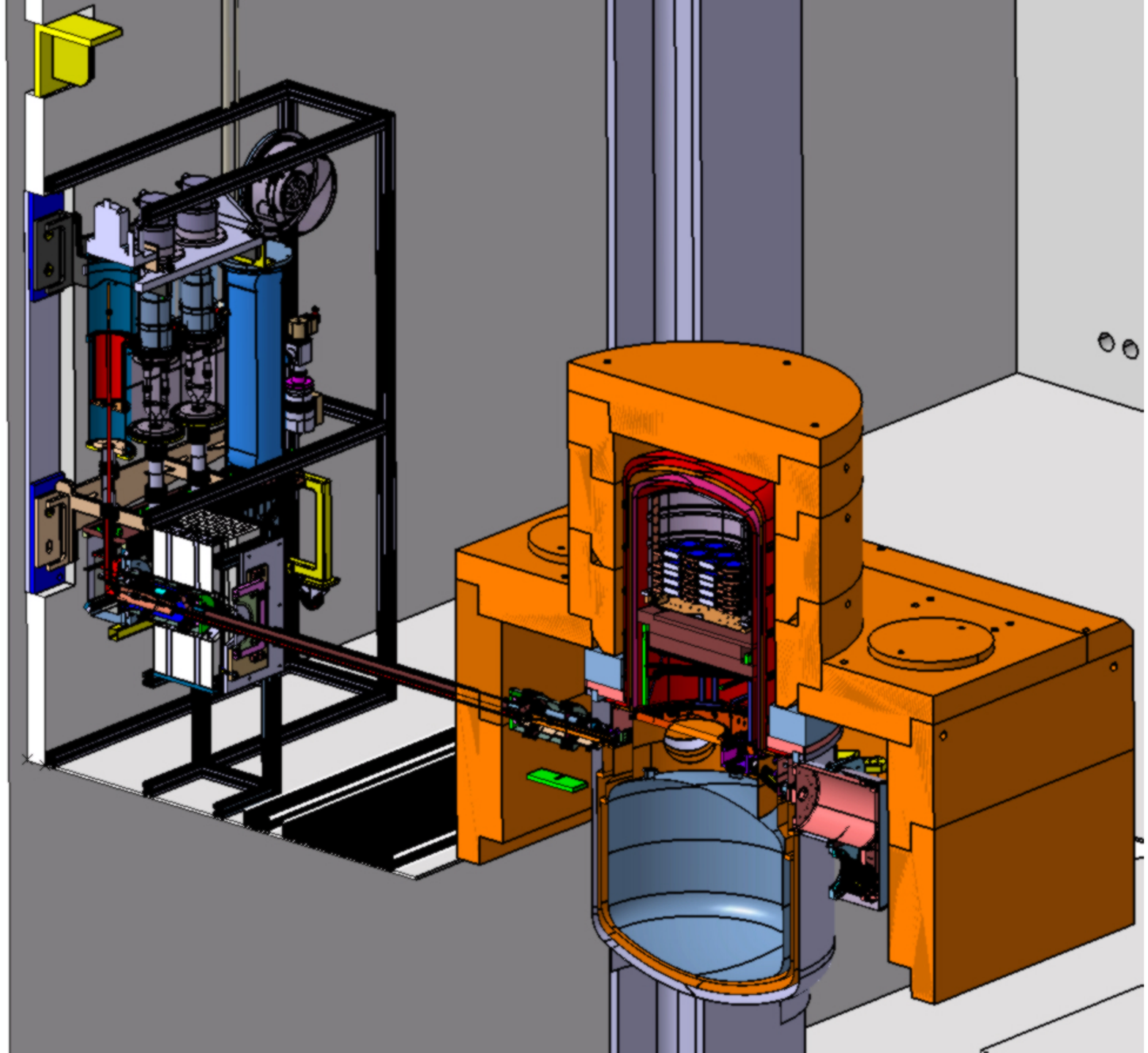}
\caption[]{Cross-section of the EDELWEISS-III reversed cryostat connected by a cryoline to the three thermal machines mounted on the cavern wall to reduce vibrations and radioactive background inside the shields. Only a  part of the lead shield is shown (orange) and the He dewar is the blue structure at the bottom.}
\label{fig:CryoEDWIII}
\end{center}
\end{figure}

The EDELWEISS-III cryostat is an upgrade of the EDELWEISS-II one which ran at the LSM from January 2006 to February 2013. It is  a reversed dilution custom-made cryostat built with selected low-radioactivity materials.\footnote{The cryostat has been designed and built by Institut N\'eel, Grenoble.} It is a $^3$He/$^4$He dilution refrigerator with an effective volume of 50 liters for the detector setup. The cryostat has been designed with the experimental volume on top of a helium dewar. In figure~\ref{fig:CryoEDWIII}  the dewar can be seen as the blue chamber in the center, the experimental volume with the thermal screens is above. Compressors and pumps are installed on the cavern wall outside the shields and connected by a cryoline with mechanical decoupling elements. The part containing the experimental chamber is mounted on a pneumatic damper system acting as a low-pass filter ($>$ 1 Hz).\footnote{EFFBE company, now Gamma-SA company, see  http://www.gamma-sa.fr.} This system is equipped with remote-controlled level adjustment capability, which is useful for optimizing resilience against mechanical interference or vibration.

\begin{figure}
\begin{center}
\includegraphics[width=1\textwidth]{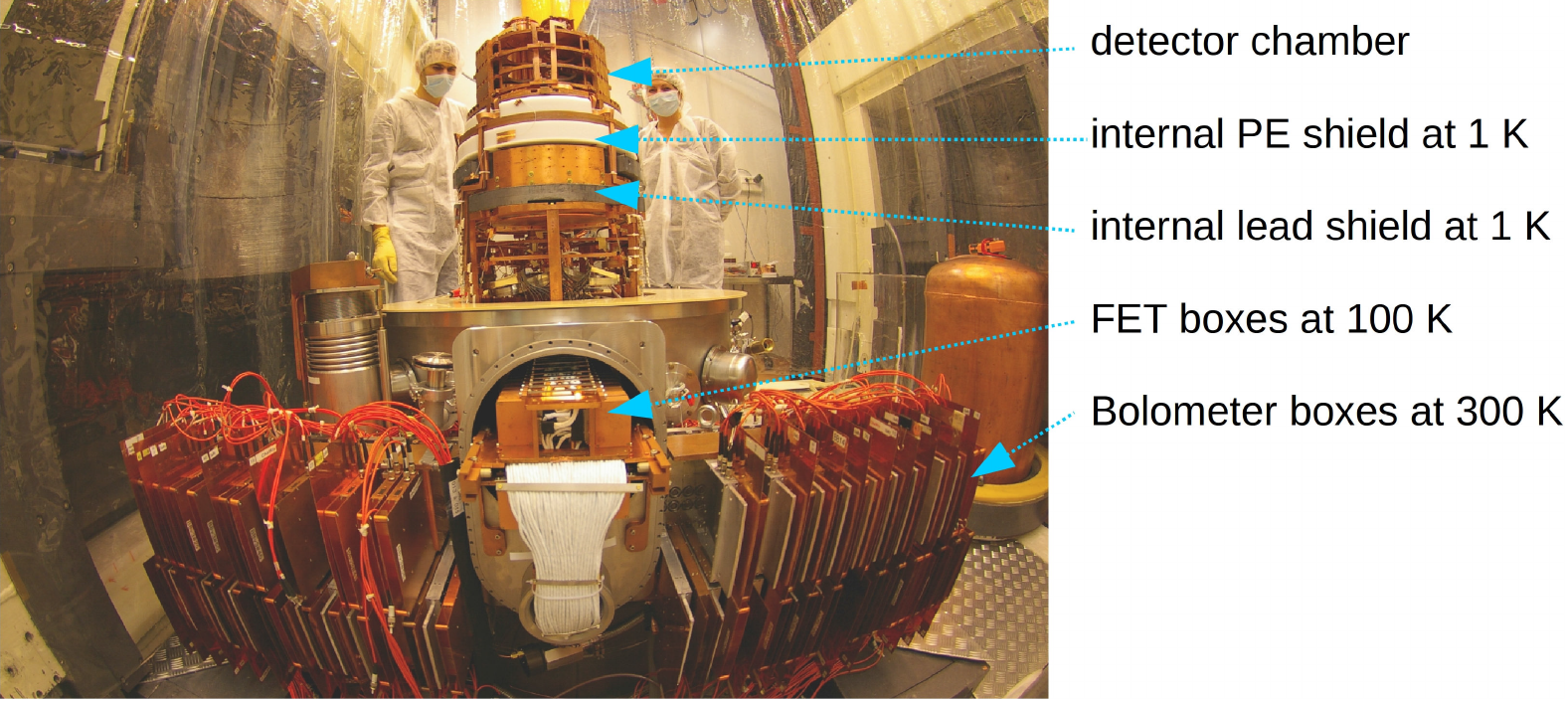}
\caption[]{Picture of the open cryostat with installed bolometer boxes and cold FET boxes. The internal lead and PE shields are visible below the detector chamber.}
\label{fig:PEshields}
\end{center}
\end{figure}

A picture of the open cryostat with mounted electronic components can be seen in figure~\ref{fig:PEshields}. The detector chamber in the top of the cryostat is made of four copper plates to hold bolometers, which are arranged in 12 towers.  This experimental volume is covered by five consecutive thermal copper screens like a matryoshka doll. The temperature decreases from the outermost screen at room temperature over four temperature stages from roughly  80--100~K, 45--55~K, 1--4.2~K towards 10--20~mK denominated the 100K, 50K, 1K and 10mK stages respectively.
The cooling power of the cryostat is 200~$\mu$W at 100~mK. The cooling power at the first, second and third temperature stage is 100~W, 20~W and 1.5~W, respectively.

The EDELWEISS-II cryostat has been modified significantly to reduce the microphonics level, to decrease the temperature of the first and second stages, to improve the ionization baseline resolutions and to decrease the helium consumption. As explained in section~\ref{sec:shields}, an additional shield of 14~cm Roman lead is installed inside the cryostat at the third stage (see figure~\ref{fig:PEshields}). Its main purpose is to shield the detectors from radioactive background of the bolometer box, the cold electronics (FET boxes at the 100K stage) and the connectors and cables at the 1K stage (see section~\ref{sec:readout}). For EDELWEISS-III the detector volume was slightly decreased to allow additional space for the installation of a PE shield. This shield is mounted between the 1K lead shield and the detectors, and has a temperature of $\sim 1$~K. Each bolometer installed inside the experimental volume is thermally connected to the 10~mK screen, which serves as a thermal bath.

Control of the cryostat during cool-down and data taking periods is carried out using three National Instruments FieldPoint (FP)
controllers~\cite{NationalInstruments}.
Labview-based user-interface software
has been developed to access, control and monitor all relevant parameters of the cryogenics~\cite{NationalInstruments}.
A TRMC2 controller manages the cryostat temperatures and bolometer heater, and communicates the data to the FieldPoint automats by TCP/IP protocol.\footnote{For more details see http://www.neel.cnrs.fr/spip.php?article149\&lang=en.} The 100K- and 50K-screens are cooled in series by the cold $^4$He gas cryoline at 2~bars, which is itself cooled by two Gifford Mac-Mahon single-stage  thermal machines. The cryoline allows a full mechanical decoupling of the cold heads from the cryostat. The temperature of the $^4$He gas at 2~bars is about 40~K at the entrance of the 50K-stage and 80~K at the output of the 100K-stage. $^4$He pressure and mass flow can be controlled to stabilize the 50K- and 100K-stages, with running temperatures at about 50--55~K and  85--90~K, respectively.

\begin{figure}
\begin{center}
\includegraphics[width=1\textwidth]{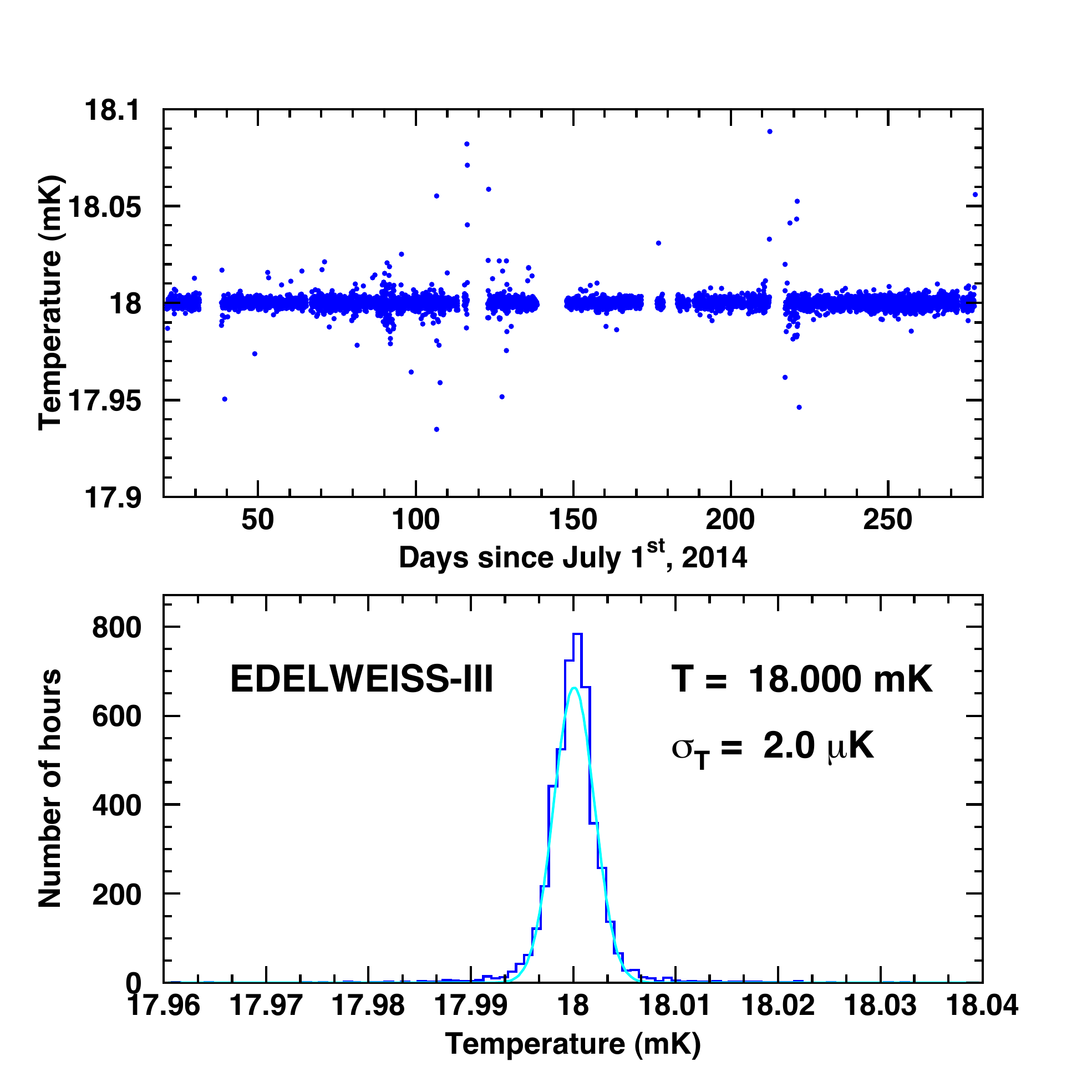}
\caption[]{Example of regulated bolometer temperature as a function of days since the start of the EDELWEISS Run 308 demonstrating the stability at 18~mK.}
\label{fig:tbolo}
\end{center}
\end{figure}
The consumption of liquid $^4$He is minimized by the use of a cold vapor reliquefaction system, based on a third Gifford Mac-Mahon machine. The whole 1K-stage is isolated from the main cryostat vacuum and is cooled by a 1K-pot controlled by a cold valve connected to the $^4$He bath.
The approximate cooldown time from room temperature to 20~mK is about one week. During standard operation, the regulated temperature is fixed at 18~mK, which takes 2--3 days more to reach. As shown in figure~\ref{fig:tbolo}, this temperature has been maintained without any major interruption during Run 308 (almost 10 months for the example shown in the top of the figure~\ref{fig:tbolo}). It is clear from inspecting the data shown in figure~\ref{fig:tbolo} that the weekly cryogenic fluid refill does not affect this temperature. The stability of the  working temperature is better demonstrated in figure~\ref{fig:tbolo} (bottom), which shows that the bolometer temperature is gaussian distributed around 18.000~mK with a variance of $\sigma_T =2~\mu$K. With this temperature the helium consumption is roughly 10 liters per day which implies a helium refill once per week.

\svnid{$Id: shields.tex 114 2017-05-23 14:52:18Z caugier $}
\section{Active and passive shielding}
\label{sec:shields}
The EDELWEISS experiment is located in the LSM~\cite{LSMwebsite} in the Frejus highway tunnel  connecting France and Italy in the Alps. The mean rock thickness of 1\,700~m (about 4\,800~m water equivalent) reduces the muon flux down to about 5~muons/m$^2$/d~\cite{muonveto}, which is more than 10$^6$ times smaller than at the surface. The natural radioactivity from laboratory walls is reduced by passive shielding, and all materials used to build the experiment as well as the detectors were selected rigorously based on their high radiopurity level. A systematic campaign of radiopurity measurements of materials was undertaken with the Gentiane HPGe detector, in operation at the LSM since 1997 and dedicated to the EDELWEISS experiment. It is an n-type HPGe diode of about 210~cm$^3$ mounted in a closed-ended coaxial configuration. Its sensitivity allows activity measurements down to the mBq level for U/Th chains, as shown in Table~\ref{table-radiopurity-measurements} of section~\ref{sec:array-performance}.

\subsection{Shielding against cosmogenic background}
\label{sec:vetomuon}

Despite the significantly reduced muon flux at LSM, cosmogenic neutrons produced by remaining muons and associated showers can still represent problematic background, if an efficient muon detection is not available (see e.g.~\cite{muonveto} and section~\ref{sec:sub:peformancemuons} of this article). An active muon veto therefore surrounds the whole experiment for an effective muon tagging~\cite{muonveto}. Fourty-six individual plastic scintillator modules (Bicron BC-412) of 65~cm width, 5~cm thickness and lengths of 2~m, 3.15~m, 3.75~m and 4~m add up to a total surface of 100~m$^2$. Each module is viewed by two groups of 2-inch PMTs, where each PMT group is provided with its individual high voltage (HV) setting and readout. In order to optimize the geometrical coverage, the muon veto has two distinct parts: the upper one attached to the external polyethylene shielding around the cryostat, and the one below the cryostat covering the pumping section of the cryostat. The upper part is positioned on rails of two movable PE blocks which allows opening of the shields right in the middle and gaining access to the cryostat during maintenance periods (figure~\ref{fig:EDW-All-Simu-Text}). The exact positions of the two mobile veto parts are monitored via regular position measurements using laser interferometry. The upper part almost completely covers the cryostat, while the lower part has a number of gaps due to cryogenic supply lines and the pillars of the experimental structure. The geometrical coverage for through-going muons is about 98\%~\cite{muonveto}. Out of a total of 46 modules, four last modules were installed in July 2010 within the upgrade program towards EDELWEISS-III. These 4 modules cover a small gap right above the cryostat appearing between the two movable parts of the muon veto. In spite of only a few cm width, this gap would lead to a significant loss in identification of coincidences between the muon veto and bolometers. The gap appeared as a result of the cryogenic upgrade, in particular, of the installation of thermal machines outside of the shielding and the corresponding cryogenic line with a diameter larger than in EDELWEISS-II (see section~\ref{sec:cryogenics}).

The muon veto DAQ is based on VME electronics and in-house developed cards. It is separate from the bolometer readout but both systems are synchronized with a 10~$\mu$s clock. The synchronous clock was upgraded for EDELWEISS-III to ensure the delivery of a reliable and precise time information (see section~\ref{sec:daq}). An event recording in the veto system is triggered once the two PMT groups of a module each pass a trigger threshold within a coincidence window of 100~ns. Within an event, all hits above threshold in the modules are recorded within a fixed interval. For each signal channel, the individual time of a hit and the integrated scintillation light are recorded. A complete description of the electronics, readout and data acquisition of the muon-veto can be found in~\cite{muonveto}.


During running periods, the status of the overall muon veto system and each of the 46 individual modules is checked online using web-based monitoring tools. They were significantly improved for EDELWEISS-III to guarantee continuous running of the system and taking data of good quality, including e.g.\ the data rate of the whole system and individual modules, ADC, TDC spectra, and their variation with time for each module.

\subsection{Shielding against radiogenic background}
\label{subsec:radiogenic-neutrons}
Radiogenic gamma and neutron backgrounds arise from natural radioactivity present in the cavern rock and concrete and in the materials of the experiment.

The residual neutron background is estimated by Monte-Carlo simulations of spontaneous fissions and $(\alpha,n)$ reactions. Neutron flux in LSM below 10 MeV is mainly due to spontaneous fission of $^{238}$U and its daughters present in the cavern and $(\alpha,n)$ reactions in light materials. The averaged $4\pi$ thermal neutron flux in the vicinity of the EDELWEISS experiment, outside the shields, has been measured to be $(3.57 \pm 0.05 (stat) \pm 0.27 (syst)) \times 10^{-6}$ neutrons/cm$^{2}$/s~\cite{EDW-Rozov2010, EDW-background}. The fast neutron flux above 1~MeV outside the shields was evaluated to be $(1.1 \pm 0.1 (stat)) \times 10^{-6}$ neutrons/cm$^{2}$/s~\cite{Rachid}. These fast neutrons can affect the sensitivity for dark matter search since they produce nuclear recoils of similar energy (few keV to few tens of keV) to those expected from WIMPs. Radiogenic neutrons are moderated and some of them are captured in the PE shielding of at least 50~cm thickness which follows the muon-veto (see figure~\ref{fig:EDW-All-Simu-Text}). The 35 tons of this PE castle reduce the fast neutron background component in the experimental volume by 5--6 orders of magnitude~\cite{EDW-background}.

The inner part closest to the cryogenic detectors is shielded using 20 cm thick lead covering the dilution refrigeration unit and the cryostat structures. The 40 tons of the Pb shielding are used against gamma backgrounds. Lead itself contains the long-lived radioactive isotope $^{210}$Pb with a half-life of 22.3 years. Therefore, the innermost 2~cm of the lead shielding are made of Roman lead from a sunken galley~\cite{muonveto} with an activity in $^{210}$Pb of less than 120~mBq/kg, more than two orders of magnitude lower than the other 18~cm of low radioactivity lead ($<$30 Bq/kg of $^{210}$Pb). 
Note that the shielding efficiency of the 2~cm Roman Pb is equivalent to 10~cm of copper.

Another significant source of background arises from the radon isotope $^{222}$Rn. This radioactive gas with a 3.8~days half-life is a decay product of $^{238}$U present in the rock and construction materials. The radon level is controlled in the whole laboratory where its measured activity is $\sim$10--15~Bq/m$^{3}$ \cite{LSMwebsite}, thanks to the ventilation system supplying air from outside of the mountain at a circulation rate of  two laboratory volumes per hour. In addition  the experiment is located in a clean room (class 10000) to avoid contamination from dust. It is mounted on the mild steel structure with rails, which allows the opening of the shields in two parts to access the cryostat. The radon level is also controlled in the EDELWEISS clean room and the empty space between the lead shield and the outermost thermal screen of the cryostat is flushed with radon depleted air with a residual activity of about 30~mBq/m$^{3}$~\cite{NEMOradon}.

Simulations performed for the EDELWEISS-II experiment showed that some specific materials presenting a high contribution to neutron background had to be replaced~\cite{EDW-background}.
\noindent The EDELWEISS-III setup was notably improved with respect to  the previous phase of the experiment. As shown in figure~\ref{fig:PEshields}, a new internal PE shield was added between the detection volume and the 1K-plate and another PE shield was added outside the cryostat above warm electronics. Furthermore, the copper used for the cryostat thermal screens, detector housings and for the 10~mK area was replaced by much purer copper (NOSV Electronic Tough Pitch copper produced by Aurubis formerly Norddeutsche Affinerie)~\cite{SilviaLRT, AURUBIS}.

\svnid{$Id: readout.tex 115 2017-05-23 17:10:10Z caugier $}
\section{Readout of signal channels}
\label{sec:readout}

The EDELWEISS-III electronics has been designed to obtain a FWHM baseline resolution below one keV on both heat and ionization signals to provide a 10$^5$ discrimination factor for nuclear recoil deposits against electron recoils above 10~keV. The corresponding electrical signals to be read out simultaneously are voltage variations over few M$\Omega$ NTD thermistors (heat) and charge measurements on the four Al  electrodes collecting electron-hole pairs (ionization).
The use of Si-JFET based cold electronics is well adapted to these high-impedance signals.\footnote{JFET is used for junction gate field-effect transistor.} Low temperature stages are available to cool parts of the electronic components, helping to reduce the current noise of the JFETs as well as thermal noise from resistances. Long integration times are possible thanks to the low event rate of only a fraction of Hz. The main challenges concern the long distance between detectors and the first elements of the cold electronics, the power dissipation allowed by the cooling power of the dilution unit and the cryostat, and finally the radiopurity requirements.

Connections from the detectors to the bottom of the detector chamber are made with in-house developed Kapton cables~\cite{these-Xiaohe-Zhang}. Beyond this point, EDELWEISS-III is using special cryogenic coaxial cables designed in collaboration with the Axon Cable group.\footnote{For more details see http://www.axon-cable.com/en/.} It is needed to keep the cross-talk between adjacent channels below few percents and to minimize the stray capacitances (40~pF/m), the sensitivity to microphonics, the electromagnetic noise and the power dissipation. The cold JFETs are mounted on extractable boxes screwed on the 100K screen of the cryostat (see figure~\ref{fig:DataStream}).

The electronics for the JFET biasing, DACs to bias the detectors, post-amplification, anti-aliasing filtering and digitization are all integrated in a single room-temperature module (so-called bolometer box), which is directly bolted onto the cryostat (see figure~\ref{fig:DataStream}). All input/output functions of this module are carried out via optical fibers. The data of all channels are digitized continuously at a rate of 100~kHz
with 16-bit resolution and sent to computers via dispatching units. Filtering, triggering and data flow control are done online in the trigger computers (see section~\ref{sec:daq}). A common control, sampling frequency and clock allows easy identification and subtraction of common noise patterns due to electronic interference.
   \begin{figure}[h]
   \begin{center}
      \includegraphics[width=1\textwidth]{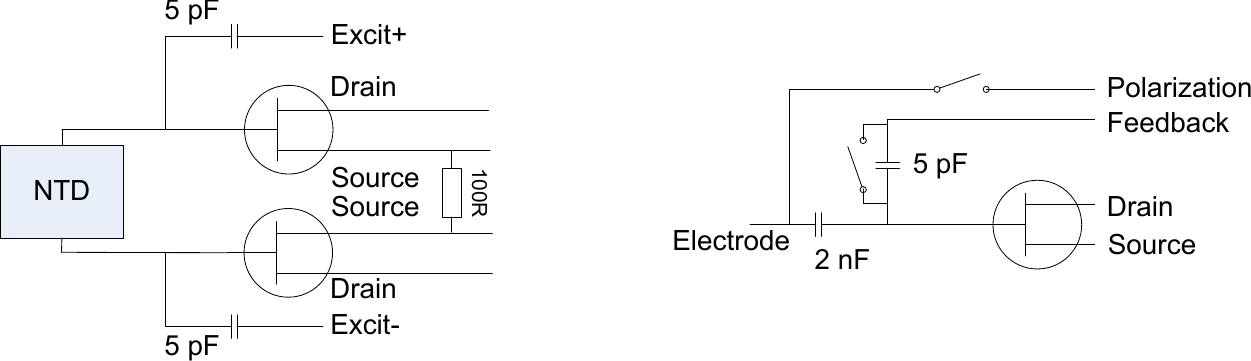}
      \caption{EDELWEISS-III signal readout: heat (left) and ionization (right) channels.}
                        \label{fig:readout}
   \end{center}
   \end{figure}

\subsection{Heat channel readout}
\label{sec:sub:heat-channel}
The experimental optimal heat energy resolution is obtained with impedan\-ces of the NTD thermistors of a few M$\Omega$ and a current bias of a few nA at 18~mK. These values are a compromise between the sensitivity to microphonics at such high impedance and non-linearity of the $V_{NTD}(I)$ responses (NTD voltage versus bias current). These non-linearities appear due to thermal decoupling of the detectors from the bath via the heat sink, thermal decoupling of the NTD from the germanium crystal via the glue in between them and electron-phonon decoupling within the NTD~\cite{billard}.

The typical heat capacity for a fully equipped FID detector is about 2~nJ/K at 18~mK, so the expected temperature elevation for an energy deposition of 1~keV is roughly 0.1~$\mu$K. With a typical heat response of 0.5~V/K, an energy sensitivity of 50~nV/keV is obtained. Typical time constants at 18~mK are 10~ms for the rise time and 10 and 100~ms for the two main decay times respectively (see~\cite{billard} for details).

EDELWEISS-III heat pre-amplification electronics is based on IFN860 bi-JFETs with a noise plateau at about 1--2~nV/$\sqrt{Hz}$ rms. The white noise contribution of the NTD Johnson noise is in the same range. Considering only these two sources of white noise and a decay time constant of 100~ms lead to an electronic baseline resolution of about 300~eV FWHM. Heat current excitation is modulated in the 500~Hz range: at both ends of the NTD two opposite current squares are created by differentiating a triangular voltage pulse on a 5~pF capacitance (figure~\ref{fig:readout}, left). This square modulation technique has been originally developed for the Archeops balloon and Planck satellite~\cite{ACreadout}. The demodulation is done after the digitization at 100~kHz either online or offline by means of software. All noise that has not been modulated is rejected and the noise level is the one at the modulation frequency. The electronics 1/f noise is thus efficiently rejected by the modulation and the common-mode noise is rejected by the differential measurement. A reference square could be subtracted before the second stage of amplification to enhance the dynamic range.

\subsection{Ionization channel readout}
\label{sec:sub:ionization-channel}
The mean energy to generate an electron-hole pair is 3~eV in germanium~\cite{3eVperpair}. Therefore a 1~keV ionization energy baseline resolution corresponds to a charge of 330 electrons equivalent at the input. If an electric field is applied through the detector, the electron-hole pairs drift through the detector before being collected on electrodes. During this charge drift of a fraction of a $\mu$s, charges are induced on the collection electrodes; these charges could be integrated on the feedback capacitance of a charge amplifier or on the total input capacitance in the case of a voltage amplifier. It can be noted that the signal-to-noise ratio is not affected by the presence of a feedback capacitance and is the same for both schemes.
EDELWEISS-III ionization electronics is based on IF1320 JFETs. The main difference with respect to the previous electronics scheme of EDELWEISS-II is the use of a voltage amplifier to measure the ionization. The first stage of amplification consists of a follower, and no detector bias and feedback resistors are used to avoid their thermal noise contribution (figure~\ref{fig:readout}, right). Based on the ADC output sampled at 100~kHz, the gate voltage is periodically adjusted on the timescale of minutes through a capacitance by firmware in an FPGA. The detector is biased through mechanical relays. The low FET and detector leakage currents ($< 1$~fA) allow biasing of a detector once every few hours, otherwise the relays are open and the detector is floating.
As a consequence, the response of the system to a charge deposit is a step function. The FID800 detector capacitance is about 150~pF, the cabling stray capacitance is about 100~pF and the FET gate to source capacitance is about 50~pF. The detector response at the input is thus 180~nV/keV.
The integration time for the charge signal is only limited by pile-up. With the adopted integration time of 1~s, we obtain a baseline resolution FHWM at the 500~eV level, a factor two better on average than those achieved in EDELWEISS-II~\cite{edw2}. The ionization resolution is limited by the current noise of the FET, introducing a 1/f noise below 1~kHz~\cite{EDWIIIelectronics}.

\begin{figure}
\begin{center}
\includegraphics[width=0.9\textwidth]{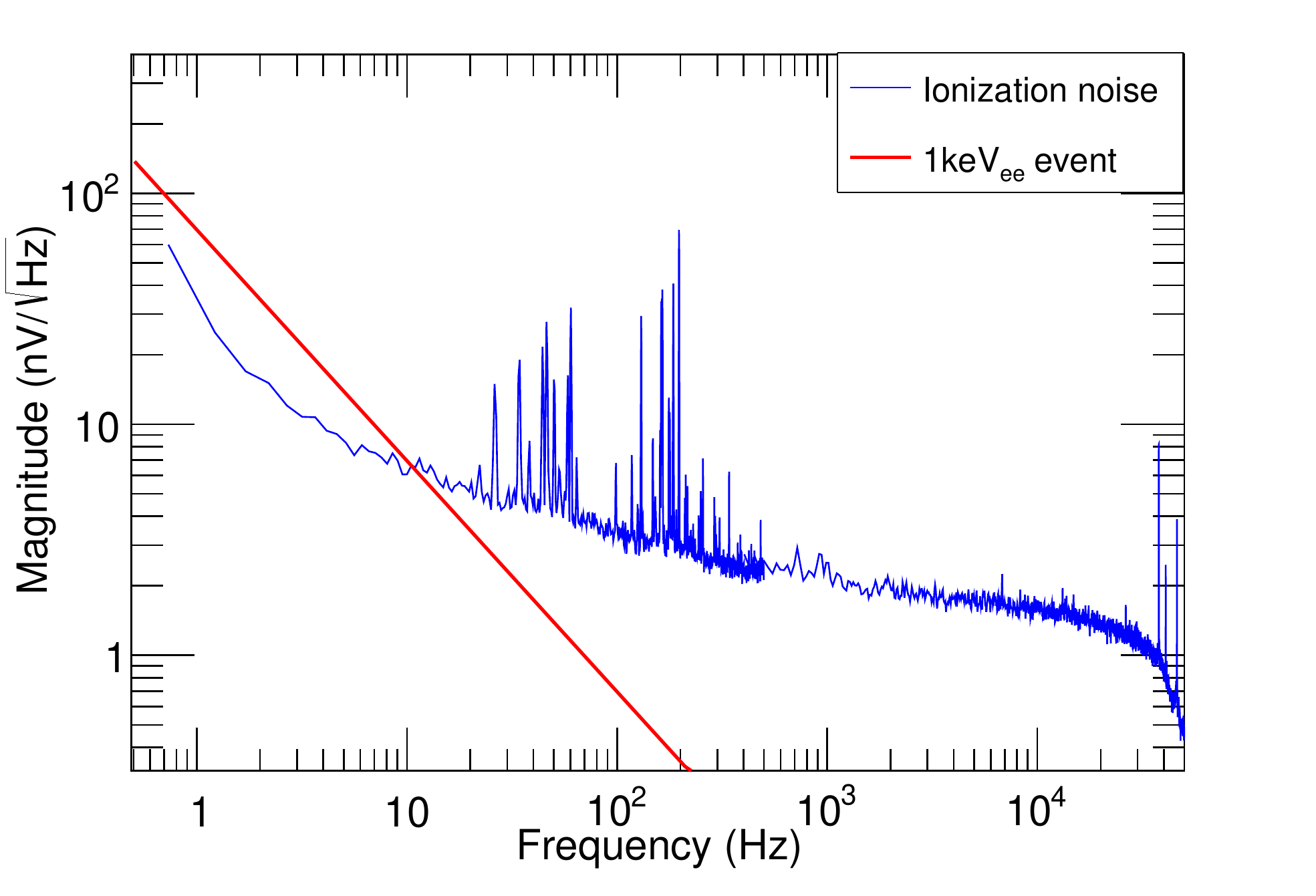} \linebreak
\linebreak
\includegraphics[width=0.9\textwidth]{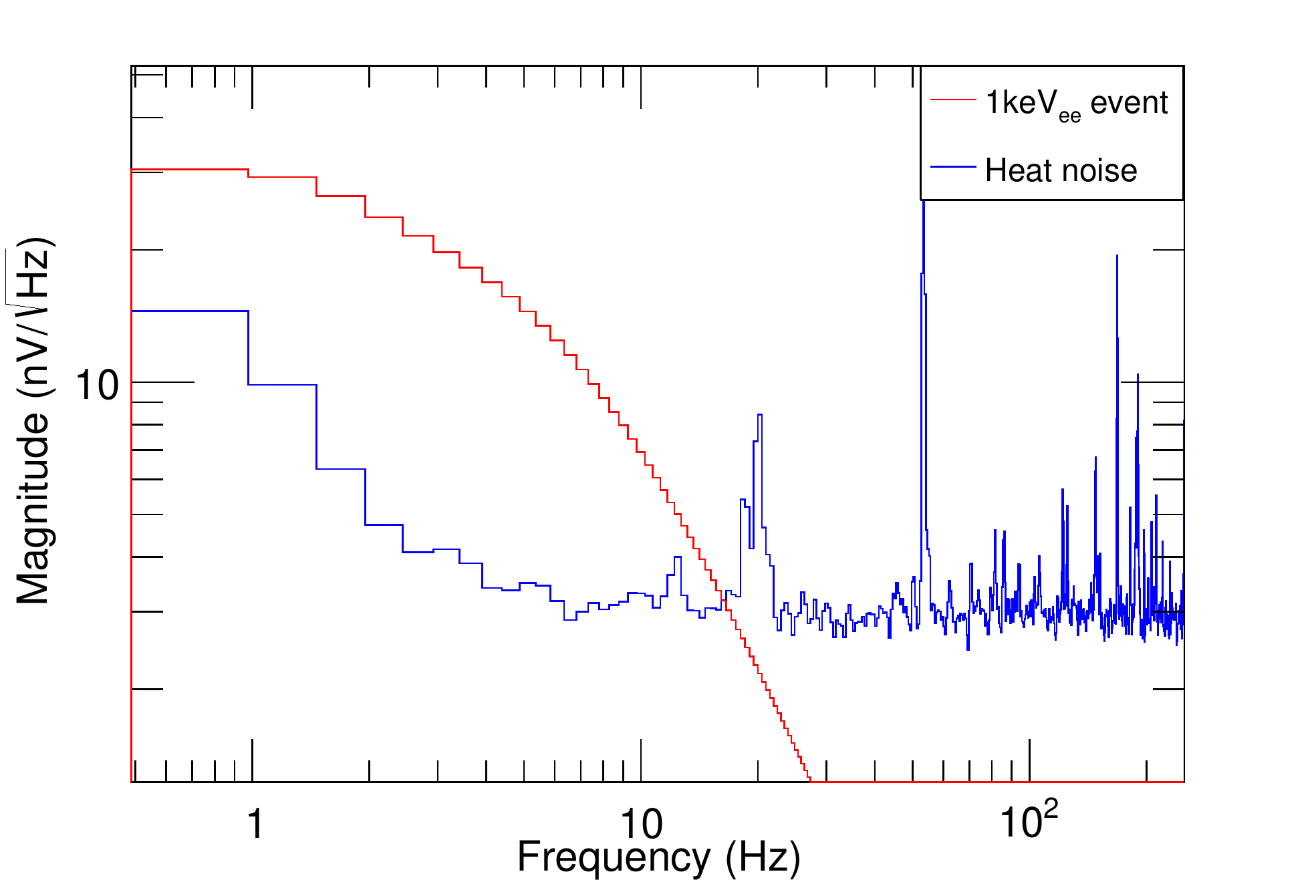}
\caption[]{Typical average Fourier noise spectra of an ionization channel (top) and a heat channel (bottom) for an FID detector. In both figures, the red histogram corresponds to the spectrum of a 1~keV$_{ee}$ (electron equivalent) signal.}
\label{fig:Noise-spectra}
\end{center}
\end{figure}

As an illustration of both ionization and heat channel readout features, figure~\ref{fig:Noise-spectra} presents noise amplitudes in the frequency domain for one FID detector. The bottom part shows the noise amplitude for one of the two heat channels whereas the top part shows it for one of the four ionization channels, after application of the amplitude correction described in~\cite{charge-paper}. This amplitude correction allows to reduce the peak structures due to the correlated microphonic noise between the four electrodes, but it also minimizes the smooth envelope mostly due to uncorrelated noise, without affecting the amplitude even in case of charge trapping. The associated power spectra expected for a 1~keV electron equivalent signal are also shown in both figures. Associated expected resolutions FWHM from an optimal filtering are 0.68~keV$_{ee}$ for the ionization channel and 0.21~keV$_{ee}$ for the heat channel shown in figure~\ref{fig:Noise-spectra}.

\subsection{Time resolved ionization readout}
It was shown in \cite{id} that a rise time analysis of an ionization signal can help to discriminate surface and fiducial events. As part of the EDELWEISS-III R\&D program, a prototype readout card allowing sampling at a rate of 40~MHz with 16~bit resolution was implemented in one of the bolometer boxes. The card receives two analog channels from one detector, hosts FPGA and 8~MB total memory, whereas 2~MB is assigned per channel and further divided into two memory pages for deadtime-free readout of up to two consecutive events. Signal splitting into the standard 100~kHz and the 40~MHz branches is done between the analog pre-amplification and anti-aliasing filters inside a bolometer box such that it does not affect the EDELWEISS standard acquisition. The digital data rate of the prototype card is 1.28~Gbit/s and exceeds the available bandwidth of 3.2~Mbit/s of the optical fiber communication between bolometer boxes and DAQ crate. Therefore, an event based readout for this type of channel~\cite{ipedaq15} is implemented through the DAQ crate and ORCA software (see section~\ref{sec:sub:orca}) to reduce maximally possible 256~kSamples per channel to $\pm$2048 points around the rising part of a pulse~\cite{thesisSiebenborn}.

\subsection{Upgrade for the Neganov-Luke amplification}
\label{subsec:luke-neganov}
As described in refs.~\cite{neganov,luke}, charges drifting under an electrical field produce additional heat in a detector by Joule effect. The measurement of such voltage-assisted heat opens access to very low ionizing events of just several 10's~eV$_{ee}$, therefore providing good prospect for low-mass WIMP searches, as shown in~\cite{cdmslite14,cdmslite16}. Within the EDELWEISS-III R\&D program, ten bolometer boxes were modified to allow biasing of ionization channels up to $\pm$70~V instead of standard $\pm$10~V, which corresponds to a maximum boost factor of about 48 compared to the non-amplified case (0~V) for electron recoils. This is done by means of operational amplifiers (LTC6090) inserted in each bias line and operated as a ``non-inverting amplifier''. The amplifier output is connected to the detector via a 20~G$\Omega$ resistor to limit the maximum current per detector in case of leakage.
The input of the amplifiers is the standard bolometer voltage of $\pm$10~V. The supply voltage of $\pm$70~V is provided by the external power supply EA~PS-2843 (max. output 84~V, 3~A).\footnote{EA-Elektro-Automatik, PS 2000 B Triple Handbook, tech. rep. (2013), http://www.\-farnell.\-com/datasheets/1634399.pdf.}
The output of the external power supply is filtered with three stages of low pass filters and currently fanned out to ten bolometer boxes~\cite{thesisSiebenborn}. The implemented scheme does not require any changes to the pre-amplification electronics and needs only minimum adaptation of the acquisition software SAMBA (see section~\ref{sec:sub:samba}).



\section{Data acquisition system}
\label{sec:daq}

\begin{figure}[h]
   \begin{center}
      \includegraphics[width=1\textwidth]{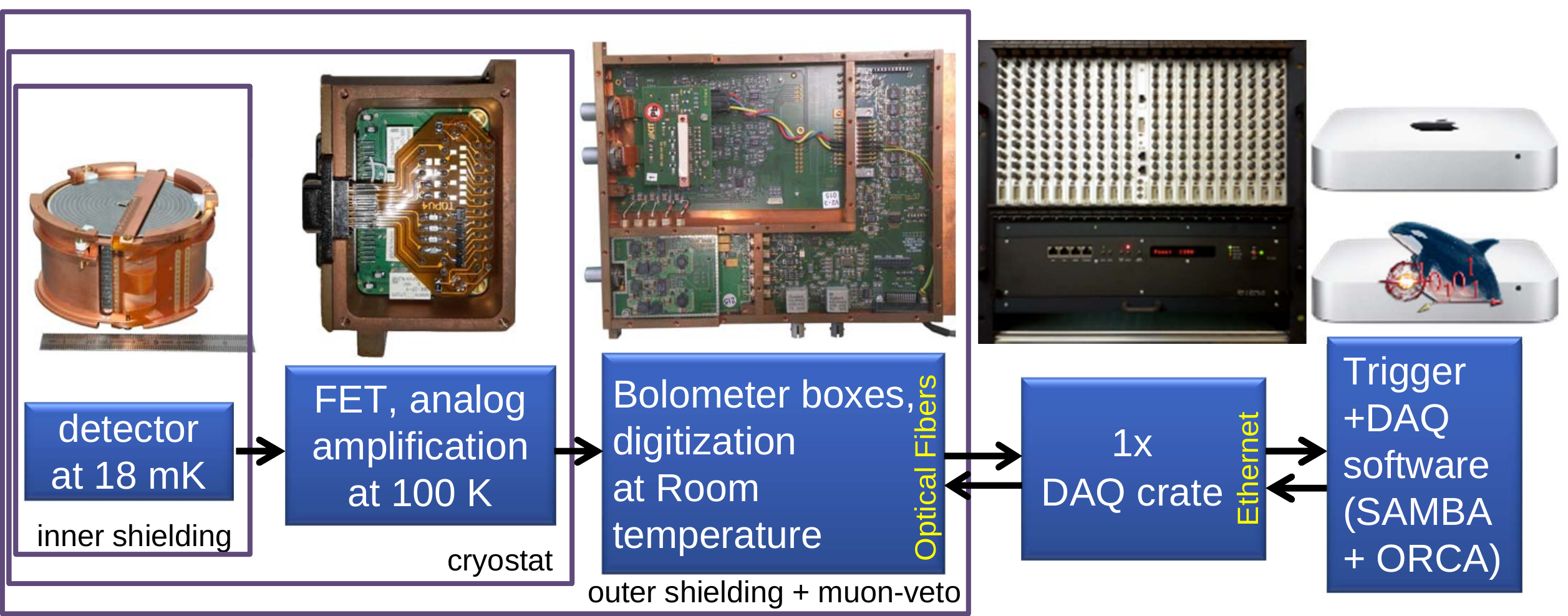}
      \caption{Full EDELWEISS-III data path. 36 detectors are read out via an analog
amplification inside the 100-K zone by 72 digitizer bolometer boxes at room temperature. Each bolometer box sends 12~Mbit/s on fiber optics to the DAQ crate. This connects via a server PC and a network to the user interface Macs. Figure adapted from~\cite{thesisSiebenborn}.}	    \label{fig:DataStream}
   \end{center}
   \end{figure}

The EDELWEISS data acquisition system (DAQ) commands the bolometer boxes (see section~\ref{sec:readout}), and therefore controls the biases of the detector sensors. It also manages the digitized data flow that they produce. It provides displays of that data for monitoring purpose, and performs triggering to select and store data on disk. The main user interface program, called SAMBA, runs on a Mac computer (right side of figure~\ref{fig:DataStream}). It receives data and sends commands via Ethernet to the DAQ crate,
consisting of a server computer connected to a VME-based acquisition crate. This system communicates with the bolometer boxes via optical fibers. The transition from analog to digital data is performed in the bolometer boxes, directly attached to the cryostat. Therefore any potential ground loops are avoided, with the ground potential of the cryostat being the same as that of the bolometer boxes.

The use of Ethernet for the input to/output from the DAQ crate enables it to communicate with more than one computer. This flexibility is used to spread the data flow and commands over multiple Mac computers, each controlling up to twelve detectors. It is also connected to an additional PC for additional monitoring and handling the fast ionization channel data flow.
The possibility to spread the data flow over a number of parallel computers provides a significant operational flexibility when operating a segmented detector array such as EDELWEISS.
The separate data sets recorded by each computer and the muon veto system are synchronized using the unique 10~$\mu$s time stamp distributed by the DAQ system.

\subsection{The DAQ crate}

The DAQ crate receives the digitized data from the optical fibers coming from the 72 bolometer
boxes controlling the 36 detectors.\footnote{For each detector, one box controls the four ionization channels and one heat channel, while the other box controls the second heat channel and the relays.} It formats the data into Ethernet packets to be sent to the DAQ computers and transmits through optical fibers the commands sent by the DAQ computers to the bolometer boxes, as well as a common clock to all boxes and the muon veto system. It also includes a minimal subset of the muon veto data into the bolometer data.

\begin{figure}[ht]
   \begin{center}
      \includegraphics[width=1\textwidth]{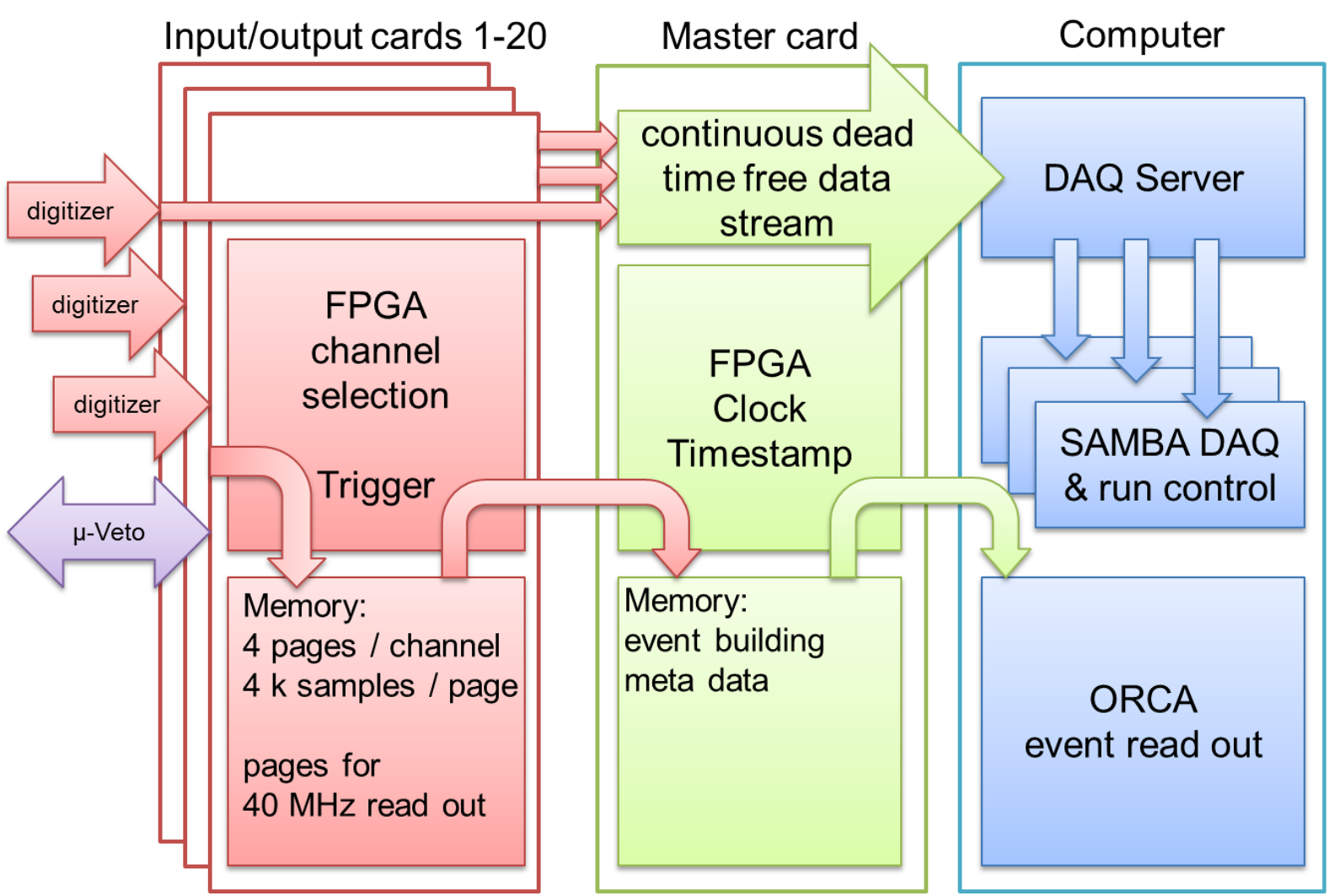}
      \caption{Schematic and simplified view of the data flow in the integrated DAQ system. The system combines an event based readout branch (ORCA, section~\ref{sec:sub:orca}) with an independent trigger logic and the dead time free data stream to the EDELWEISS computers (SAMBA, section~\ref{sec:sub:samba}). Figure extracted from~\cite{thesisSiebenborn}.}
   \label{fig:dataflow}
   \end{center}
   \end{figure}

The system has been designed as an integrated system scalable for $\geq$~40 detectors, able to deliver a reliable time synchronization. It features data-reduction capabilities to optimize the network traffic by adding or removing channels depending on their use. The DAQ crate has been developed and manufactured in-house at KIT~\cite{ipedaq12}. It consists of a 6U+2U high 19-inch crate, housing a custom backplane allowing to connect 20 input-output cards (IO-cards) and one central master card. This type of DAQ crate is also used in other astroparticle physics experiments such as the Pierre Auger observatory~\cite{auger15} and KATRIN~\cite{katrin-tdr04}. For the EDELWEISS experiment the DAQ series have been adapted to the optical fiber input/output to the bolometer boxes. Each IO-card has 6 optical fiber input-output pairs to manage up to 6~bolometer boxes, or 3 detectors. The IO-cards receive the digitized data streams from the bolometer boxes, concentrate it, and remove unused channels. One IO-card receives from the muon veto system a six-bit pattern corresponding to an event in one of its six faces, as well as time stamp information to check the synchronization of the system. In addition, the IO-cards can be programmed to perform event-based readout, a feature used to read out  the fast ionization channel in parallel to the 100-kHz data stream. The data stream from the IO-cards is forwarded to a master card in the VME crate
(see figure~\ref{fig:dataflow}). In addition to providing the 48-bit 100-kHz clock used to synchronize the flow of data and commands, the master card sends the data stream via a PCI-Express interface to a Linux-based server computer. The server then produces UDP (User Datagram Protocol) packets that are transmitted to the acquisition computers via a Gigabit Ethernet network.

With 6~channels per detector (4~ionization and 2~heat channels) sampled at 100~kHz with 16-bit resolution, each detector produces a 1.2~MB/s stream of physical data. The total throughput of 43~MB/s for 36 detectors is well below the values of up to 90~MB/s that the DAQ crate has been proven to handle~\cite{ipedaq12}.

The three Mac computers each handling the 100-kHz data flow from up to 12 detectors
are running a program that can also send the commands needed to configure the
bolometer boxes (see section~\ref{sec:sub:samba}). The commands sent to the bolometer boxes follow the reverse path: from the acquisition computer to the server via UDP packets, through the VME cards via the PCI-Express interface and then back to the bolometer boxes via different optical fibers. A separate Mac computer provides the user interface to handle the data acquisition of the fast 40-MHz ionization data (see section~\ref{sec:sub:orca}).

The 100-kHz clock and the corresponding time stamp generated by the master card is not only used to synchronize the sampling of the ADC on all channels, but also provides the frequency and phase of the NTD excitation (section~\ref{sec:sub:heat-channel}) and the 64-second cycle for the gate correction of the ionization readout (section~\ref{sec:sub:ionization-channel}). The time stamp information is used to remove the transient associated with these patterns and, in the case of the gate correction, to calculate precisely the corresponding dead time.

        \subsection{SAMBA}
 \label{sec:sub:samba}

The application used to manage the EDELWEISS DAQ hardware configuration and the data taking is called SAMBA. The program has been designed for flexibility: for example, the three computers
handling up to 12 detectors each can be operated independently, or under the control of a single master computer. The setup is defined in text files that can be modified either directly or via the SAMBA application itself via graphical interfaces. The hardware setup that SAMBA must handle includes, among others, lists of detectors to be read out, their associated bolometer boxes and associated DAQ crate channels, the bias values for the ionization channels, the frequency and amplitude of the heat excitations, and the software-controlled gains of the digitizers. Other parameters refer to the processing of the data flow, with software filters, optimized individually for each of the channels, and to the configuration of the trigger that controls the recording of data on disk. Through object-oriented programming, SAMBA can be readily adapted to detector and hardware evolution.

SAMBA receives from the crate server the 100-kHz data from up to 12 detectors. Two levels of digital processing can be performed on each channel. The first corresponds to on-the-fly filters applied to the data before it is stored on disk. A copy of that stream, not saved on disk, can be further processed in order to be used for triggering purpose. The data at either levels of processing can be displayed continuously, either in a time series (in an ``oscilloscope'' mode), or as power spectra as a function of frequency. The available digital filters are the demodulation of the triangular excitation of the NTD (section~\ref{sec:sub:heat-channel}), digital IIR (Infinite Impulse Response) filters, averages, the removal of a repetitive pattern, and the convolution with a user-defined pulse template. Different filters can be applied to each of the channels.

In addition to the ``oscilloscope'' mode, SAMBA can trigger on a single filtered channel: for instance, during WIMP-search runs the trigger is set on heat detector channels. The second-level filtered data is inspected and, if this channel exceeds a threshold value, a time tag is registered. SAMBA then compares the time tags of all channels to look for coincidences inside a given time window, in which case it determines a common time tag. For each time tag, the first-level data of all channels of the corresponding detectors are saved on disk. The data saved on disk are typically 2.048-seconds wide traces of the heat and ionization signal, sampled at 0.5 or 1~kHz and centered on the time tag, as well as 40.96-ms wide traces of the raw 100-kHz ionization data.\footnote{For the oversampled ionization, each point is the average of hundred or two hundred 100-kHz samples. For the heat, each point is the difference between the average signal during the positive and negative phase of the NTD excitation.}
The data of neighboring detectors is generally saved in addition, in order to provide
ample baseline data for noise and trigger studies. Each event is accompanied by a header indicating the bit pattern of the triggered channels, the level of the trigger threshold for each channel, and the amplitude of each channel in the same units as the threshold value.

SAMBA can be programmed to automatically adjust the trigger level on a given channel in order to
maintain a given trigger rate. For instance, in WIMP-search runs, it is typically required that
each channel triggers not more than three times per minute and less than thirty times every 10~minutes. If this is not the case, the trigger threshold level is automatically raised, or decreased, respectively, by a fixed amount. This results in a rather constant triggering rate of 50~mHz per channel, irrespective of slow fluctuations in noise levels. As physical event rates in WIMP-search runs are close to 5~mHz per detector, this procedure ensures lowest possible thresholds without being burdened by excessive dead time.

\subsection{ORCA}
\label{sec:sub:orca}
An alternative interface to the DAQ crate is given via an acquisition computer running the ORCA
program. This interface, which can configure the DAQ crate, is used in particular when exploiting
the 40~MHz time resolved ionization channels. ORCA stands for Object-oriented Real-time Control and Acquisition and is written for the Mac OSX operating system in Objective-C~\cite{snodaq04}.
It was originally provided with the DAQ crate system, and was adapted to the EDELWEISS data format, the different FPGA register model for hardware access and the bolometer access. In case of 40~MHz time resolved channels, the acquisition computers cannot forward the software triggers of SAMBA fast enough to send a trigger signal to the bolometer boxes and request the 40~MHz ADC traces. Instead, FPGA triggers on the crate IO-cards must be used~\cite{ipedaq15}. Once an event is detected, the ADC trace centered on the event is written to a buffer on the FPGA. Otherwise, the ADC data is discarded. The ORCA readout loop of the DAQ crate computer polls the hardware, reads out events from the buffers and sends the ADC traces to ORCA after adding additional information like time stamps, channel and crate number. ORCA stores the events in a run file, which can be later converted into ROOT~\cite{root} format by means of OrcaROOT library.
The time-resolved data can be merged with the SAMBA events using the
associated time stamp data. As the SAMBA and ORCA acquisition systems are relatively independent, the comparison of the event lists they produce is useful to study potential acquisition dead time~\cite{thesisSiebenborn}.

\svnid{$Id: performance.tex 116 2017-05-24 11:58:21Z caugier $}
\section{Detector performance}
\label{sec:performance}

The use of the newly designed 800-g FID detectors coupled with upgrades on both electronics and cryogenics systems led to the improvement of the average FWHM of the heat and ionization baseline energy resolutions and better active rejection of background as described in this section.

The double measurement of the heat and ionization energies provides both the recoil energy  $E_r$ deposited by a particle interacting in the detector and the recoil type via the estimation of the ionization yield $Q(E_r)$ as a function of the recoil energy~\cite{recoil-energy}:
\begin{equation}
E_r = E_{heat} \left( 1 + \frac{V}{\varepsilon_{\gamma}} \right) - E_{ion} \frac{V}{\varepsilon_{\gamma}} \,\,\, {\rm and}  \,\,\, Q(E_r) =  \frac{E_{ion}}{E_r}
\label{Er-Q-1}
\end{equation}
where $V$ is the bias voltage and $\varepsilon_{\gamma} = 3$~eV 
is the average energy needed for an electron recoil to produce an electron-hole pair of charge carriers in germanium. $E_{heat}$ and $E_{ion}$ stand for the heat and ionization energies, respectively, obtained from signal amplitudes, calibrated using electron recoils. The quenching variable $Q(E_r)$ is equal to 1 for electronic interactions by definition, whereas $Q(E_r) \sim 0.3$ for nuclear recoils~\cite{recoil-energy}. Thus the simultaneous measurement of heat and ionization provides an event-by-event identification of the recoil type and allows rejection of the dominant $\gamma$-ray backgrounds as well as the majority of $\beta$-backgrounds. Residual $\gamma$- and surface $\beta$-backgrounds are removed using the active rejection power of the FID detector design, as presented in sections~\ref{sec:gamma-rejection} and \ref{sec:surfacerejection}.

\subsection{Energy calibration and detector regeneration procedures}
\label{sec:Calib}

Due to the presence of impurities inside the detector, charge carriers can be trapped by these impurities during their migration towards the electrodes. They then cause the formation of space charges that create an electric field in a direction opposite to that from the bias voltage. To limit this effect, so-called regeneration phases are conducted for one hour per day, where $\gamma$-radiation is used to break the formed space charges.

Energy calibration and regeneration procedures make use of radioactive sources and have been carried out for all detectors. To avoid background due to these sources during science data taking, they have to remain outside the EDELWEISS-III shielding during standard data acquisition and have to be put as close as possible to the detectors only during calibration and regeneration procedures. For this purpose, an automatic source deployment mechanical system has been designed and implemented. This system incorporates two $^{60}$Co and two $^{133}$Ba radioactive sources.

\subsubsection{Radioactive sources}

For the detector regeneration, strong $^{60}$Co $\gamma$-sources ($\sim 200$~kBq) are used. Calibration of ionization and heat signals of the detectors is carried out using simultaneously two encapsulated sources of $^{133}$Ba.  The radioactive isotope $^{133}$Ba decays by electron capture to $^{133}$Cs ($T_{1/2}$ = 10.55~yr), with the emission of $\gamma$-rays among which the relevant ones for calibration are
356.0 keV (62.05\%) and 383.9 keV (8.94\%)~\cite{nudat2}, which have enough energy to reach the bolometers inside the cryostat.

\begin{figure}[ht]
\begin{center}
\includegraphics[width=1\textwidth]{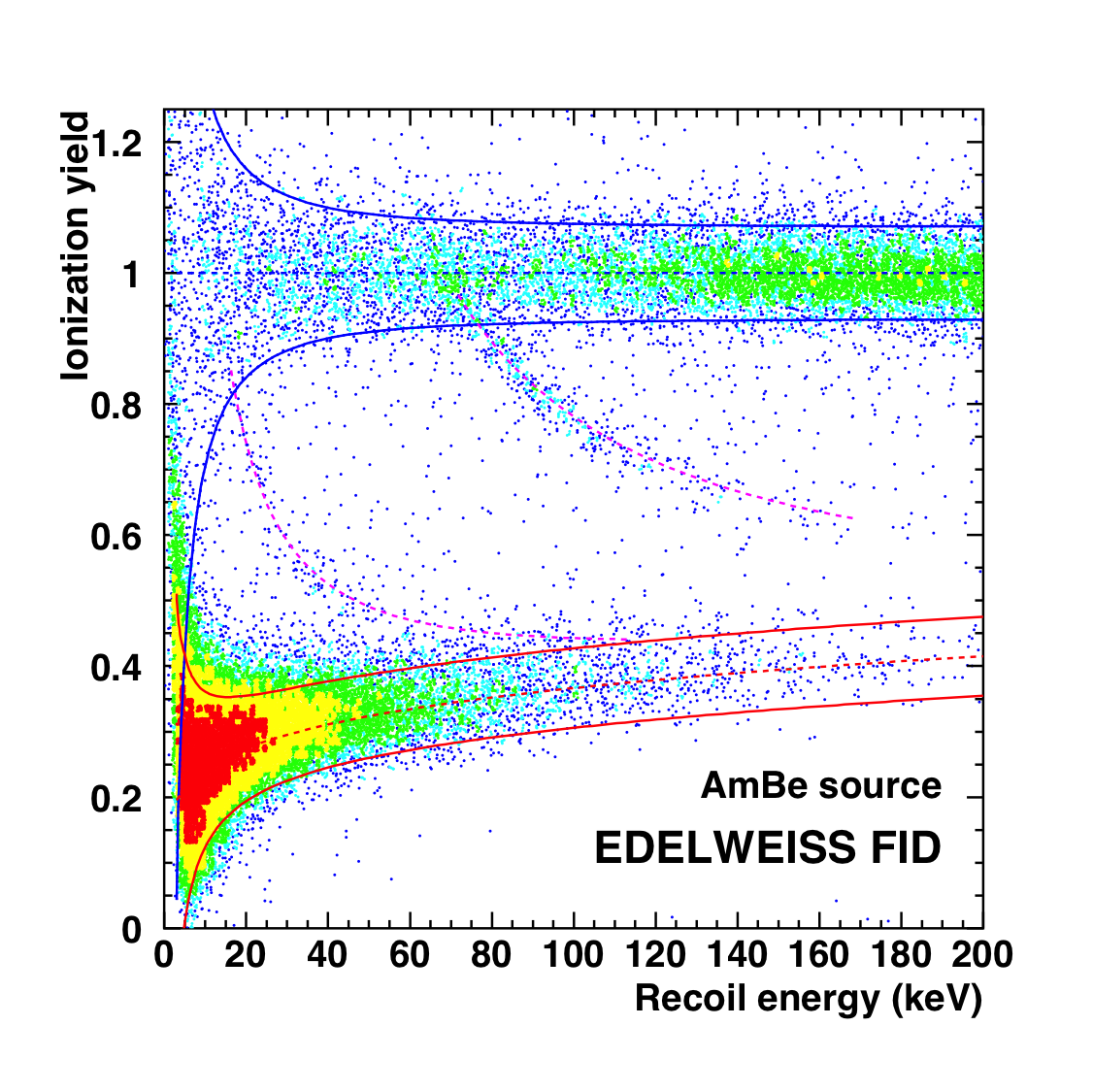}
\caption[]{Ionization yield versus recoil energy for a large statistics ($> 3 \times 10^4$) of events from a neutron calibration using an AmBe source. The two red (blue) solid lines delimit the 90\%\,C.L.\ nuclear (electron) recoil band.
Purple dashed lines correspond to inelastic scattering of neutrons on the first (13.28~keV) or the third (68.75~keV) excited state of $^{73}$Ge.}
\label{fig:AmBe-calib}
\end{center}
\end{figure}

A low activity ($21 \pm 4$ neutrons/s)
AmBe neutron source is used for neutron calibrations. The source is encapsulated in lead and sealed in a stainless steel container. For the purpose of neutron calibration the source is manually placed at a distance of 8~cm above the top of the cryostat inside the Pb shielding for several hours. Figure~\ref{fig:AmBe-calib} shows the ionization yield $Q(E_r)$ versus the recoil energy $E_r$ for AmBe neutron calibration data. The source emits neutrons with energies up to 11~MeV, inducing nuclear recoils with $Q(E_r)$ values gaussian-distributed around a smooth function $Q(E_r) = 0.16 E_r^{0.18}$ consistent with the expectations from the Lindhard theory~\cite{Lindhard} including effects due to multiple scatterings~\cite{Q-Qprime}.

The AmBe source also emits high energy $\gamma$-rays of 4.4~MeV which lose energy via Compton scattering, leading to the population of events distributed around $< Q(E_r) > = 1$. Events between the electron and nuclear recoil bands arise from the inelastic scattering of a neutron on a germanium nucleus. The events along the purple dashed lines correspond to those involving the first (13.28~keV) or third (68.75~keV) excited state of $^{73}$Ge.\footnote{The 66.7 keV line from the second excited state (half-life of 0.499~s) should appear in the electron recoil band, but it is invisible since it is rejected during the analysis as being similar to an afterpulse.} The prompt $\gamma$-emission results in an effective quenching $Q_{eff}$ which is the average of the $Q$ values of the neutron and the $\gamma$, weighted by their energies.

\subsubsection{Energy calibration procedure}
\label{sec:Ionization-heat-recoil-energies}
\label{sec:sub:Calib}
\label{Er_and_Q}

The calibration of ionization channels is performed using the 356~keV peak of the $^{133}$Ba spectrum to adjust individual gains and cross-talk coefficients. Data taking dedicated to $\gamma$ calibrations represents around 10\% of the data over a cool-down period.

As explained in~\cite{charge-paper}, energy losses are correlated with the energy deposit depth, with a dependence always stronger for the fiducial electrode $B$ (anode) than for the fiducial electrode $D$ (cathode), since the charge trapping is significantly larger for electrons than for holes. 
The resolution increases with the energy of the signal, with a linear term which is dominated by charge trapping effects. Thus the degradation of resolution with energy is a consequence of charge carrier trapping in the bulk of the detectors.\footnote{As demonstrated in~\cite{charge-paper}, event dispersion in the electron recoil band is correlated with charge trapping. It allows correction of trapping effects and improvement of the energy resolution of both fiducial channels.
This charge trapping correction procedure also improves heat energy resolution.
Nevertheless, no clear improvement is achieved below $E_r = 50$~keV. For this reason the energy correction is not applied to WIMP search analysis as we expect measured ionization energies due to a WIMP interaction to be below 50~keV.}
It follows that despite good fiducial ionization baseline resolutions between 450 and 600 eV (FWHM) for most of the 800-g FID detectors, as presented in section~\ref{sec:baseline-resolutions}, measured resolutions at 356~keV are as large as $5\% \times E_{ion}$. Nevertheless, these resolutions are sufficient to determine the absolute energy scale with a satisfactory precision. The procedure is carried out in two steps: the fiducial electrode $D$ is first used to adjust the peak position; then the ratios of the amplitude of the $D$ electrode to all other electrodes are determined. The gains and the cross-talk coefficients are obtained for all the electrodes.

After the calibration procedure, individual ionization energies $E_{ion-x}$ ($x=a,~b,~c,~d$) are reconstructed for both fiducial electrodes $B$ and $D$ 
and both veto electrodes $A$ and $C$ of each FID detector. Both electronics cross-talks and gains are observed to be constant throughout an overall cool-down period.
Due to the FID detector electrode implementation, cross-talk coefficients associated with adjacent channels are around 20--30\%.

The heat gains are stable over periods of months but may vary within a few percent between two short data taking periods (around 24h) or at the beginning of a cool-down. These gains are measured using the heat-over-ionization ratio of $\gamma$-ray background events during WIMP search and calibration runs. This ratio is also used to measure and correct the non-linearities of the heat channels as a function of energy. After calibration one obtains the linear heat energy $E_{heat-1}$ $E_{heat-2}$) for NTD1 (NTD2), for which electron recoils with energy $E_r$ in the fiducial volume have $E_r = E_{fid} = E_{heat-x}$, with $x =1,~2$  (see Table~\ref{tab:var-def} for definitions of variables).

\subsubsection{Definition of energy variables}
\label{sec:energy-variables}

After the calibration procedures, all energy scales are expressed in keV electron equivalent (keV$_{ee}$). For the purpose of the analysis, individual ionization and heat energies can be combined into other energy variables, as seen in eqs.~\eqref{energy-combinations1} and \eqref{energy-combinations2}:
\begin{equation}
E_{dif} = \frac{(E_{ion-b}-E_{ion-d})}{2} \,\,\,\, ; \,\,\,\, E_{fid} = \frac{(E_{ion-b}+E_{ion-d})}{2}
\label{energy-combinations1}
\end{equation}
\begin{equation}
E_i = \frac{(E_{ion-a}+ E_{ion-b}+ E_{ion-c}+E_{ion-d})}{2} \,\,\,\, ; \,\,\,\, E_{veto} = E_{ion-a}+ E_{ion-c}
\label{energy-combinations2}
\end{equation}
where 
$E_{dif}$ is the difference between the charge collection measured on the two fiducial electrodes,
$E_{fid}$ is the fiducial ionization energy, $E_i$ is the total ionization energy
and $E_{veto}$ is the veto energy used to define surface events with respect to bulk events. \\
In addition, $E_c$ is defined as the combined heat energy, which corresponds to the resolution-weighted average of the measured heat energies $E_{heat-1}$ and $E_{heat-2}$ of the two NTDs. All the variables associated with energies and baseline resolutions
are listed in Table~\ref{tab:var-def}.\footnote{Resolutions FWHM of channel or channel combination $var$ are defined as FWHM$_{var} = 2.35 \times \sigma(E_{var})$.}
\begin{table}[ht]
\begin{center}
\caption{Energy variables and the associated baseline resolutions.}
\begin{tabular}{|c|c|}
\hline
Variable & Description \\ \hline
$E_{ion-a}$, $E_{ion-c}$  & Ionization energy of veto channel $A$ and $C$ \\
FWHM$_{ion-a}$, FWHM$_{ion-c}$ & Associated baseline resolutions (FWHM) \\ \hline
$E_{ion-b}$, $E_{ion-d}$   & Ionization energy of fiducial channel $B$ and $D$ \\
FWHM$_{ion-b}$, FWHM$_{ion-d}$ & Associated baseline resolutions (FWHM) \\ \hline
$E_{dif}$  & Difference between charge collections \\
 & measured on the 2 fiducial electrodes \\
FWHM$_{dif}$ & Associated baseline resolution (FWHM) \\ \hline
$E_{fid}$  & Fiducial ionization energy \\
FWHM$_{fid}$ & Associated baseline resolution (FWHM) \\ \hline
$E_{i}$  & Total ionization energy \\
FWHM$_{i}$ & Associated baseline resolution (FWHM) \\ \hline
$E_{veto}$  & Veto ionization energy \\
FWHM$_{veto}$ & Associated baseline resolution (FWHM) \\ \hline
$E_{heat-1}$, $E_{heat-2}$  & Heat energy of NTD1 and NTD2 \\
FWHM$_{heat-1}$, FWHM$_{heat-2}$ & Associated baseline resolutions (FWHM) \\ \hline
$E_c$  & Combined heat energy \\
FWHM$_{c}$ & Associated baseline resolution (FWHM) \\ \hline
\end{tabular}
\label{tab:var-def}
\end{center}
\end{table}

The general form of the recoil energy $E_r$ as introduced in eq.~\eqref{Er-Q-1} can be obtained for any interaction from the combined heat signal $E_c$, after an appropriate subtraction of the Neganov-Luke effect:
\begin{equation}
E_r = E_{c} \left( 1 + \frac{V}{\varepsilon_{\gamma}} \right) - \sum_{j} \left( \frac{V_j}{\varepsilon_{\gamma}} \, E_{ion-j} \right)
\label{Er-Q-1-general}
\end{equation}
where 
$V_j$ is the voltage applied on electrode $j$ and $E_{ion-j}$ the asssociated ionization energy. Note that eq.~\eqref{Er-Q-1-general} holds for interactions taking place in any part of the detector including multiple scatters.

In terms of variables $E_{fid}$, $E_{veto}$ and $E_c$, and considering that charges are collected not only on the fiducial electrodes with a difference of potential $V_{fid} = |V_{BD}|$, but also in the surface volume between one fiducial and one veto electrode with a difference of potential $V' = |V_{AB}| = |V_{CD}|$, the definition of the recoil energy $E_r$ becomes:\footnote{For bulk events with $E_{veto} = 0$, recoil energy as expressed in eq.~\eqref{Er-2} is identical as the one given e.g.\ in ref.~\cite{Sanglard}. Interestingly eq.~\eqref{Er-2} is also valid when the charge is collected only between $A $ and $C$ veto electrodes with $E_{veto} = 2E_{i}$.}
\begin{equation}
E_r = E_{c} \left( 1 + \frac{V}{\varepsilon_{\gamma}} \right) - \frac{1}{\varepsilon_{\gamma}}
\left( V_{fid} \, \left( E_{fid} - \frac{1}{2}  E_{veto}  \right) + V' \, E_{veto}  \right)
\label{Er-2}
\end{equation}
The associated ionization yield is given by:
\begin{equation}
Q(E_r) =  \frac{E_i}{E_r}
\label{Q-2}
\end{equation}

\begin{figure}
\begin{center}
\includegraphics[width=1\textwidth]{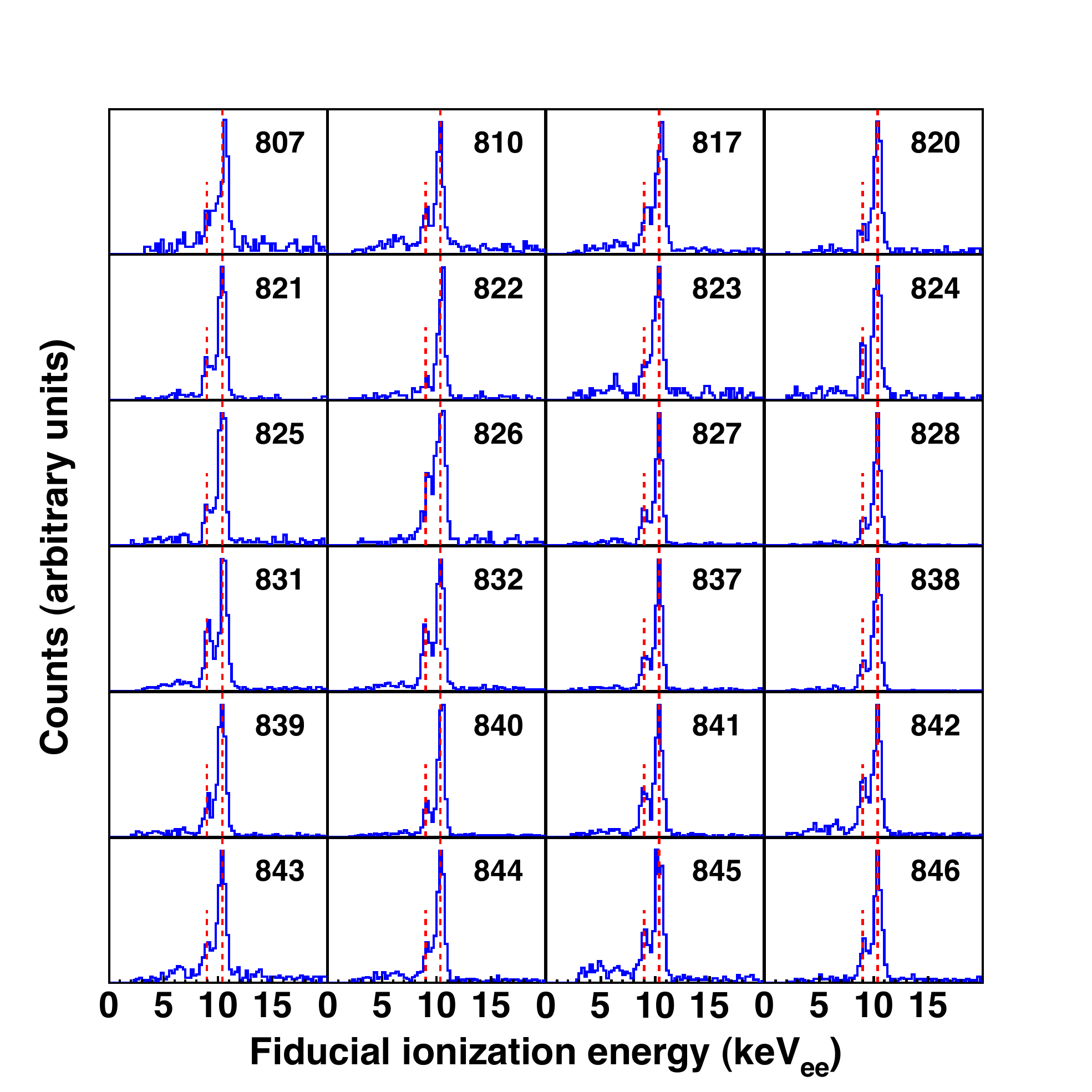}
\caption[]{Calibration of ionization energy scales: cosmogenic peaks (as described in section~\ref{sec:sub:fiducialvolume}) for the 24 detectors used for coincidence studies, measured with fiducial ionization energy $E_{fid}$. The 8.98 keV and 10.37 keV peaks are found at expected positions, as demonstrated with the two red dashed lines. 
Calibrations are well under control with non linearities lower than 5\%.}
\label{fig:24-10keV-ion}
\end{center}
\end{figure}

\begin{figure}
\begin{center}
\includegraphics[width=1\textwidth]{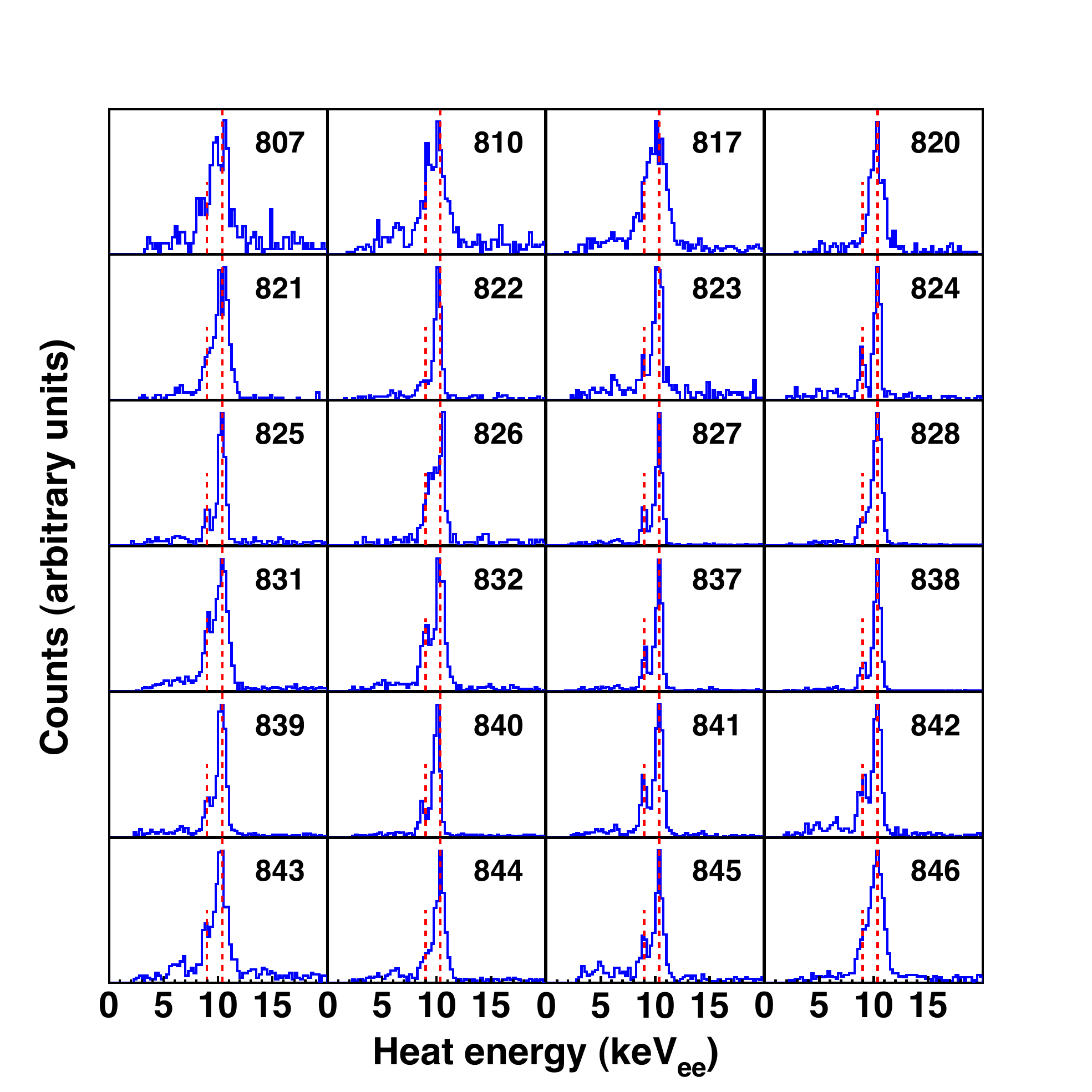}
\caption[]{Same as figure~\ref{fig:24-10keV-ion} for heat energy channels, using combined heat energy $E_c$.}
\label{fig:24-10keV-heat}
\end{center}
\end{figure}

In conclusion it is worth mentioning that despite the well-understood charge trapping mechanism leading to broadening of detector resolutions, both ionization and heat calibrations are well controlled, within 5\%, for all FID detectors. As an illustration, figures~\ref{fig:24-10keV-ion} and~\ref{fig:24-10keV-heat} show cosmogenic lines (as defined in section~\ref{sec:sub:fiducialvolume}) for the 24 detectors used in Run 308 for coincidence studies, measured with fiducial ionization energy $E_{fid}$ and combined heat energy $E_c$, respectively.\footnote{During Run 308 only 24 detectors were read out over the 36 installed in the cryostat.} In both figures the two red dashed lines correspond to the expected 8.98 and 10.37~keV cosmogenic peak positions.

\subsection{800-g FID detector baseline resolutions}
\label{sec:baseline-resolutions}

The baseline resolutions per detector are determined using events for which the considered detector did not trigger the data acquisition. The resulting signal amplitudes for each channel have gaussian distributions. Each channel resolution is determined as the standard deviation of the gaussian fit to the distribution of baseline amplitudes recorded during a time interval of one hour and leads to associated FWHM values. Average FWHM baseline resolutions and online thresholds for physical events in WIMP runs obtained during the long data taking of Run 308 are listed in Table~\ref{tab:baseline-resol-threshold} for the 24 detectors used for coincidence studies in EDELWEISS-III analyses.

\begin{figure}
\begin{center}
\includegraphics[width=0.63\textwidth]{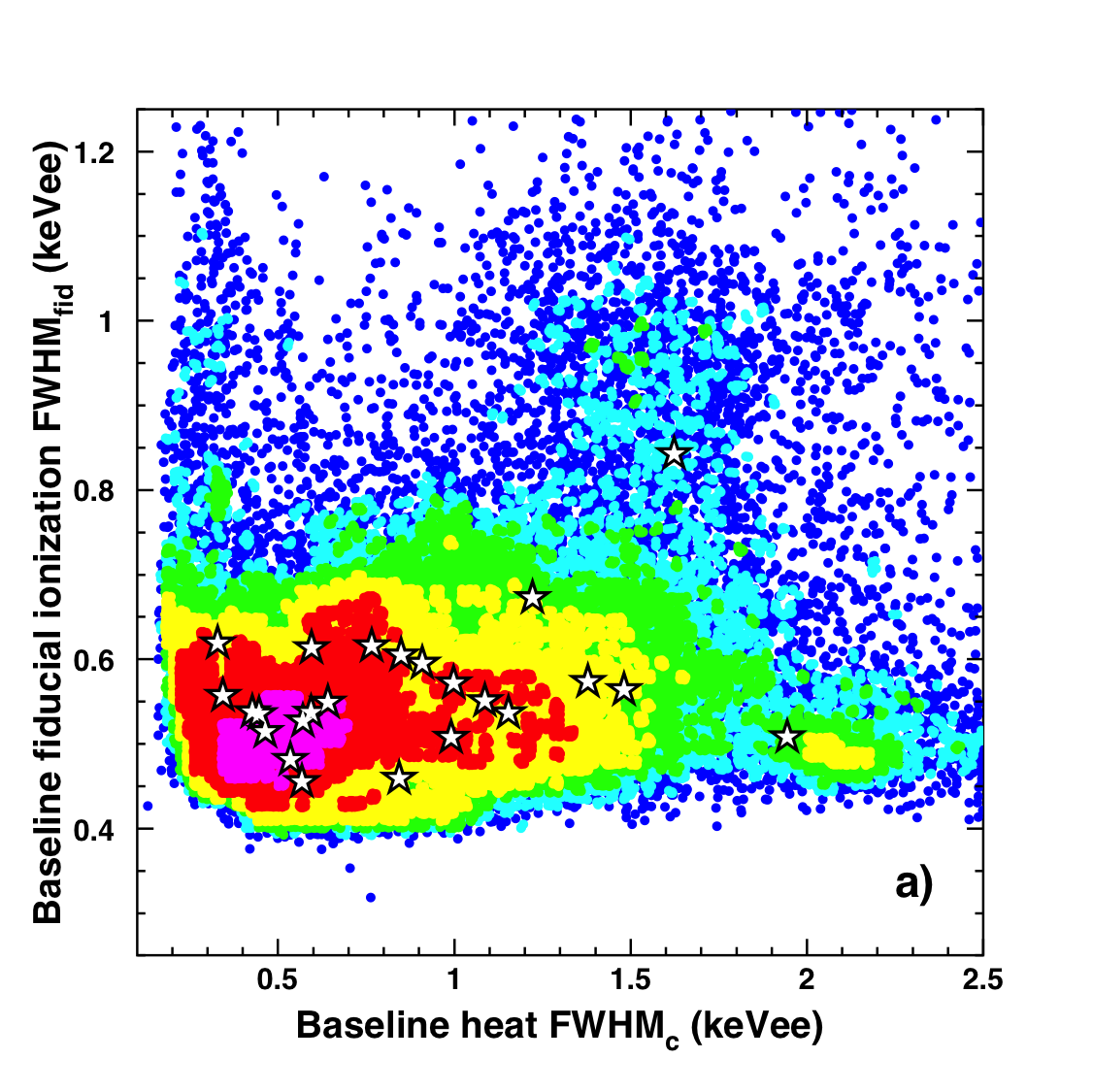}
\includegraphics[width=0.63\textwidth]{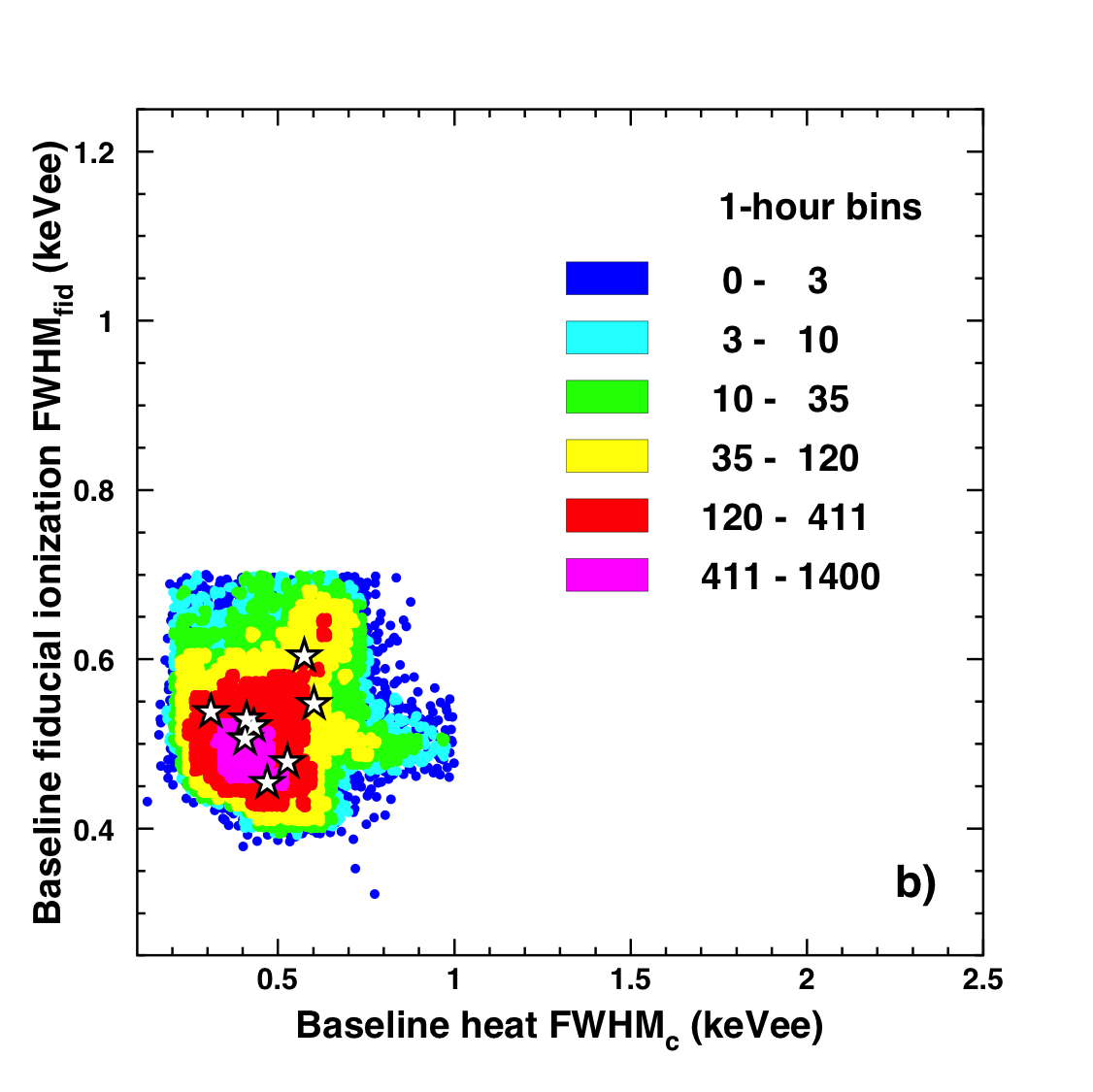}
\caption[]{Baseline fiducial ionization FWHM resolution (FWHM$_{fid}$) versus the baseline combined heat FWHM resolution (FWHM$_c$), in keV$_{ee}$ (color code denotes number of 1~h periods, stars represent average bolometer values as given in Table~\ref{tab:baseline-resol-threshold}): a) for the 24 detectors used for coincidence studies; b) after time period selection for the 8 bolometers used for low-mass WIMP studies~\cite{low-mass-paper}, with online threshold less than 1.5~keV$_{ee}$, and rejecting a small fraction of time when FWHM$_c$ is larger than 1~keV$_{ee}$ or FWHM$_{fid}$ is larger than 0.7~keV$_{ee}$.}
\label{fig:biplotreso}
\end{center}
\end{figure}

Dispersions from these average FWHM values can be estimated from figure~\ref{fig:biplotreso}~a), which presents baseline fiducial ionization FWHM resolution versus heat baseline FWHM resolution
for the 24 detectors used for coincidence studies. The color encoding shows FWHM resolution combinations occurring a certain number of 1~h periods, stars denote the average values per bolometer as given in Table~\ref{tab:baseline-resol-threshold}. Figure~\ref{fig:biplotreso} underlines that FWHM$_{fid}$ values are mainly below 0.7~keV$_{ee}$ for fiducial ionization energies whereas  FWHM$_c$ values are mainly below 1.5~keV$_{ee}$ for heat energies.
Such baseline resolutions are good enough for standard mass WIMP search using the 800-g FID detectors, while for low mass studies~\cite{low-mass-paper} a subset of 8 bolometers with clearly better resolutions has been used. In the latter analysis time-periods with online threshold less than 1.5~keV$_{ee}$ are selected, and a small fraction of time when the combined heat baseline FWHM$_{c}$ is larger than 1~keV$_{ee}$ or the fiducial ionization FWHM$_{fid}$ is larger than 0.7~keV$_{ee}$ is rejected (see figure~\ref{fig:biplotreso}~b) and in~\cite{low-mass-paper}). This tighter selection changes the resolutions by less than 10\%.

\begin{table}[h]
\begin{center}
\caption{\label{tab:baseline-resol-threshold}Average FWHM resolutions and thresholds for physical events in WIMP runs obtained during EDELWEISS-III data taking (Run 308). FWHM$_{fid}$, FWHM$_{i}$ and FWHM$_{c}$ are associated respectively with $E_{fid}$, $E_i$ and $E_c$ as listed in Table~\ref{tab:var-def}. The online threshold is the lowest of the two heat channels and FWHM$_{c}$ corresponds to the combined heat baseline FWHM resolution.}
\begin{tabular}{|c|c|c|c|c|c|c|}
\hline
 \multicolumn{1}{|c|}{Detector} &\multicolumn{1}{|c|}{FID Bias} & \multicolumn{3}{|c|}{Baseline resolutions} &  \multicolumn{1}{|c|}{Online} \\
\multicolumn{1}{|c|}{} & \multicolumn{1}{|c|}{$V_{fid}$ (V)} & \multicolumn{3}{|c|}{(keV$_{ee}$) (FWHM)} &  \multicolumn{1}{|c|}{threshold} \\
\cline{3-5}
& & FWHM$_{fid}$ & FWHM$_i$ & FWHM$_c$ & (keV$_{ee}$)\\ \hline
FID807 &        6.4& 0.84 & 1.19 & 1.62 & 3.41 \\ \hline
FID810 &        8& 0.51 & 0.78 & 1.85 & 2.93 \\ \hline
FID817 &        6.4& 0.56 & 0.85 & 1.44 & 3.41 \\ \hline
FID820 &        8& 0.57 & 0.90 & 1.03 & 2.29 \\ \hline
FID821 &        8& 0.55 & 0.89 & 1.08 & 2.39 \\ \hline
FID822 &        8 & 0.53 & 0.64 & 0.59 & 1.32 \\ \hline
FID823 &        8& 0.46 & 0.73 & 0.78 & 1.76 \\ \hline
FID824 &        8& 0.55  & 0.75 & 0.32 & 0.62 \\ \hline
FID825 &        8 & 0.45 & 0.69 & 0.49 & 0.94  \\ \hline
FID826 &         16& 0.61 & 0.90 & 0.28 & 0.47 \\ \hline
FID827 &        8 & 0.51  & 0.80 & 0.42  & 0.81  \\ \hline
FID828 &        8 & 0.59 & 0.83 & 0.84 & 1.59 \\ \hline
FID831 &        8 & 0.53 & 0.92 & 1.15 & 3.31 \\ \hline
FID832 &        8 & 0.60 & 0.89 & 0.78 & 1.65 \\ \hline
FID837 &        8& 0.54  & 0.92 & 0.43  & 0.98  \\ \hline
FID838 &        8 & 0.53  & 0.78 & 0.44  & 0.93  \\ \hline
FID839 &        8 & 0.55 & 0.92 & 0.62  & 1.20  \\ \hline
FID840 &        8 & 0.53 & 1.17 & 0.57 & 1.44 \\ \hline
FID841 &        8 & 0.48  & 0.80 & 0.53  & 0.99  \\ \hline
FID842 &        8 & 0.62 & 0.93 & 0.60  & 1.25  \\ \hline
FID843 &        8& 0.56 & 0.83 & 1.37 & 2.59 \\ \hline
FID844 &        8& 0.51 & 0.71 & 0.94 & 1.67 \\ \hline
FID845 &        8& 0.61 & 1.27 & 0.76 & 1.64 \\ \hline
FID846 &        8& 0.67 & 0.96 & 1.22 & 2.82 \\ \hline
\end{tabular}
\end{center}
\end{table}

\subsection{Fiducial volume of the 800-g FID detectors}
\label{sec:fiducialvolume}
\subsubsection{Definition of fiducial cuts}
\label{sec:fiducialcuts}
Following an energy deposit in the bulk, electrons and holes created drift towards the fiducial electrodes $B$ and $D$ whereas for surface events charge collection is shared between one fiducial electrode and one veto electrode (either $B\&A$ or $D\&C$). Fiducial cuts are used to reject surface events ($^{206}$Pb recoils or surface $\beta$ and $\gamma$ events), which could be otherwise misidentified as nuclear recoils due to an incomplete charge collection. The fiducial selection is performed requiring an equal charge of opposite sign on fiducial electrodes and no charge collected on the veto electrodes $A$ and $C$.

\begin{figure}
\begin{center}
\includegraphics[width=1\textwidth]{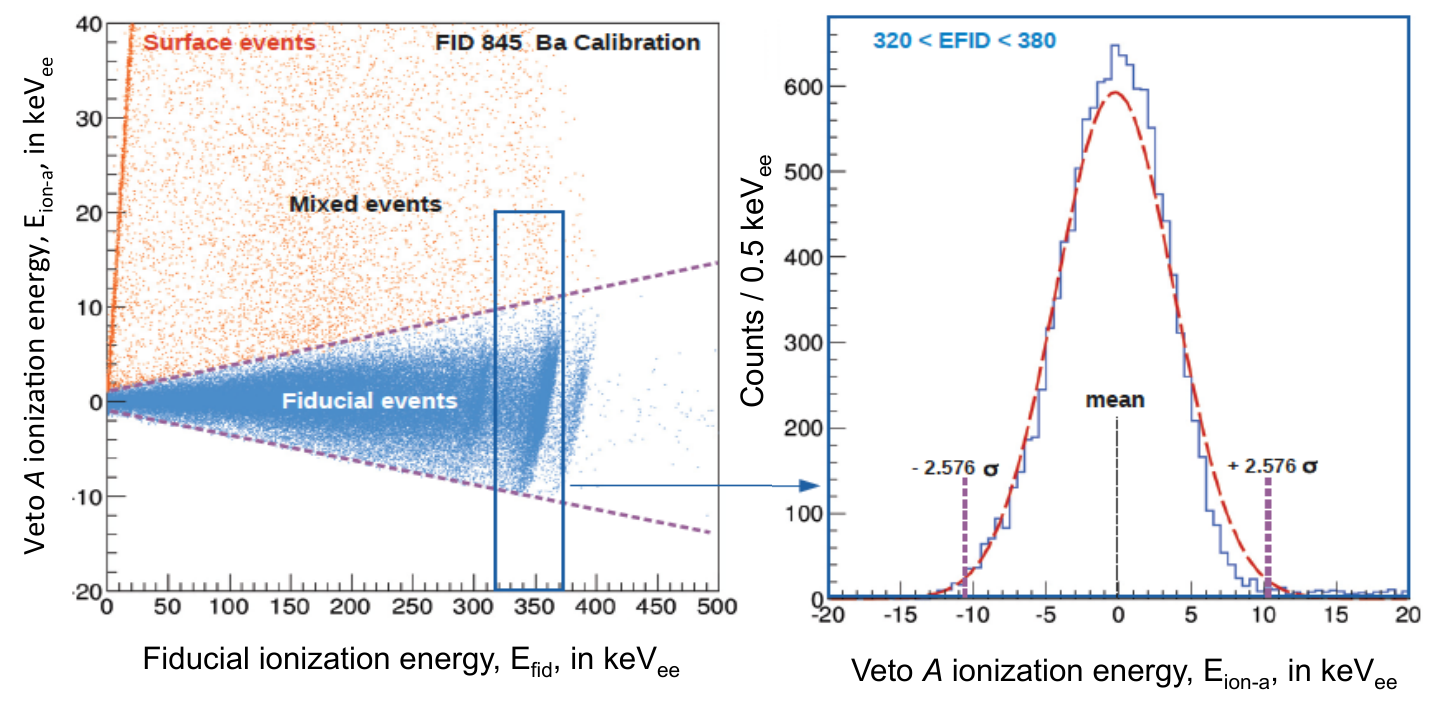}
\caption{Example of fiducial event selection. Left: distribution of events in the plane $(E_{fid},~E_{ion-a})$. Selected events after fiducial cut (dashed purple lines) are represented in blue. Overdensity of orange dots corresponds to pure surface events, other orange dots are intermediate events between the two categories. Right: energy distribution for veto  electrode $A$, corresponding to events with fiducial energy $E_{fid}$ between 320 and 380~keV$_{ee}$, fitted with a gaussian distribution of standard deviation $\sigma$. Dashed purple lines correspond to values of $E_{ion-a}$ at $\pm 2.576 \sigma$ from the adjusted distribution center. These values are used to determine the fiducial cut with 99\% efficiency for bulk events. Figure extracted from~\cite{QArnaud-thesis}.}
\label{fig:4.13Quentin}
\end{center}
\end{figure}
Three cuts are performed to optimize the surface event rejection while keeping an efficiency of fiducial event selection of at least 99\% for each cut: two of them concern the ionization energy measured by each of the veto electrodes ($E_{ion-a}$ and $E_{ion-c}$); the third cut is performed on charge collection difference $E_{dif}$. An example of the cuts is given in figure~\ref{fig:4.13Quentin}~(left), which shows the different event categories in the plane $(E_{fid}, \, E_{ion-a})$ and the associated fiducial cut on veto $A$.

The distribution of fiducial events is also defined by the energy-dependent dispersion, as shown for veto $A$ in figure~\ref{fig:4.13Quentin}~(right) for events with energies between 320 and 380~keV$_{ee}$ (main $\gamma$-line at 356~keV for $^{133}$Ba decay, see figure~\ref{fig:4.13Quentin}~(left)). This dispersion has a nearly gaussian distribution, thus using a cut at $\pm 2.576 \sigma$ from the distribution's center provides a fiducial event selection efficiency of 99\%. Due to the dispersion enlargement with $E_{fid}$, the selection has to be reproduced for different fiducial ionization energy intervals.

\begin{figure}
\begin{center}
\includegraphics[width=1\textwidth]{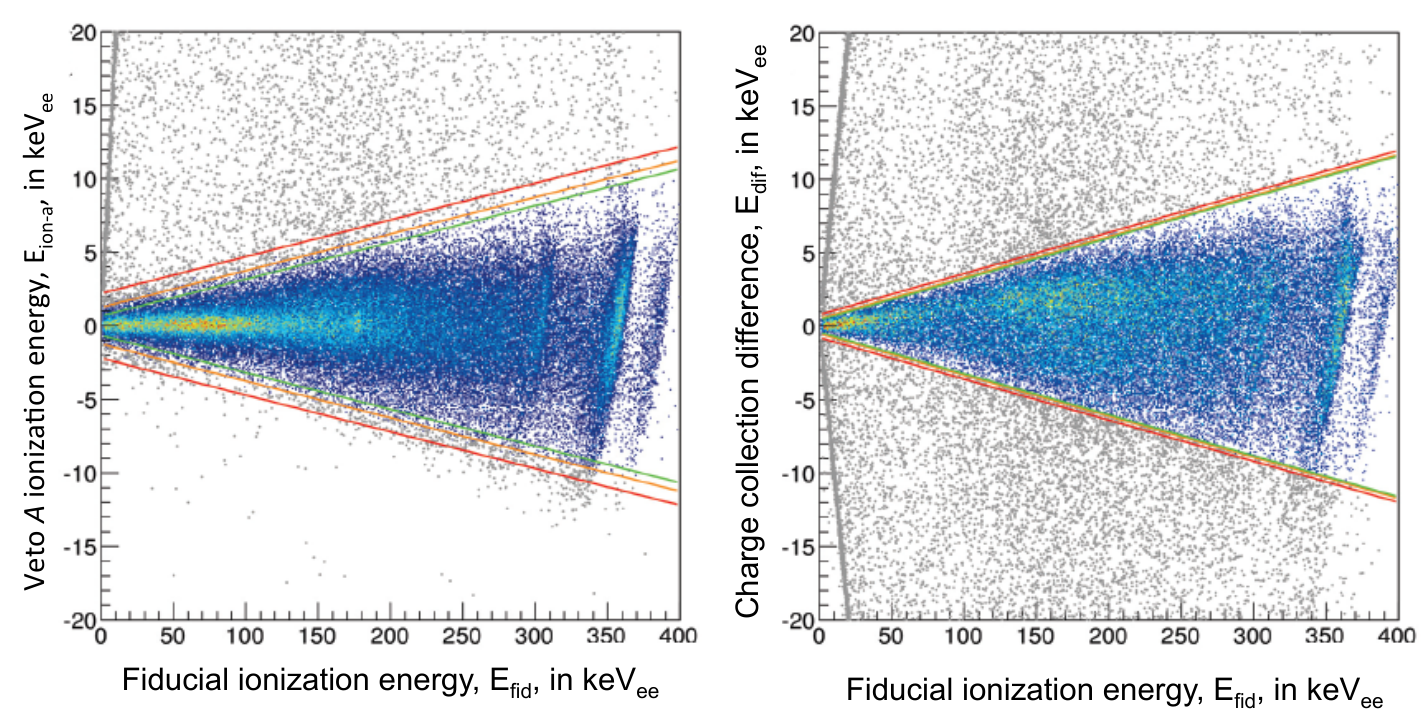}
\caption{Example of fiducial cut (orange solid lines) after optimized data selection on an hourly basis. Fiducial events are represented as color dots whereas non fiducial events are shown as gray dots. Green and red lines correspond to the fiducial cut for having the best and the worst resolution respectively. Left: cut on $E_{ion-a}$ for veto $A$ electrode. Right: cut on charge collection difference $E_{dif}$. Figure extracted from~\cite{QArnaud-thesis}.}
\label{fig:4.16Quentin}
\end{center}
\end{figure}

Fiducial cuts corresponding to baseline resolution conditions of each channel are obtained in a similar way, as shown in figure~\ref{fig:4.16Quentin} for veto $A$ (left) and charge collection difference (right) of one detector. Green (red) lines correspond to fiducial cuts during a period of one hour with having the best (the worst) resolution. Orange lines correspond to fiducial cuts for average resolutions. Surface event rejection at low energy is obtained applying quality cuts on FWHM$_{ion-a}$, FWHM$_{ion-c}$ and FWHM$_{dif}$ baseline resolutions (as defined in Table~\ref{tab:var-def}). A semi-automatic procedure is carried out using 3 conditions on $E_{fid}$ combined with $E_{ion-a}$, $E_{ion-c}$ and $E_{dif}$, as well as on their baseline resolutions~\cite{QArnaud-thesis}. It allows to ensure a fiducial selection efficiency of 99\% independent of the experimental conditions.

\subsubsection{Determination of the fiducial volume}
\label{sec:sub:fiducialvolume}

EDELWEISS-III crystals have been moved to the underground location, they have been exposed to cosmic rays at the surface. This resulted in a production of $^{68}$Ge ($T_{1/2}$ = 270.8 d) and $^{65}$Zn ($T_{1/2}$ = 244.3 d) isotopes  by cosmogenic activation predominantly due to neutron flux. These cosmogenic isotopes decay by electron capture to $^{68}$Ga ($T_{1/2}$ = 67.7 min, $E_K$ = 10.37~keV) and $^{65}$Cu (stable, $E_K$ = 8.98~keV), accompanied by an X-ray cascade measured as a single-site event inside the detectors, with energy depending on the considered element.\footnote{$E_K$ are binding energies of K-shell electrons. Values of $E_K$ energies are taken from~\cite{TOI}.} $^{68}$Ga further decays to $^{68}$Zn by electron capture (stable, $E_K$ = 9.66~keV). In addition thermal neutrons absorbed during AmBe calibration by $^{70}$Ge isotopes in the germanium crystals produce $^{71}$Ge ($T_{1/2}$ = 11.4 d), which also decays by electron capture towards $^{71}$Ga (stable, $E_K$ = 10.37~keV). More details on cosmogenic activation in EDELWEISS germanium detectors can be found in refs.~\cite{tritiumpaper, low-mass-paper}.

To determine the fiducial volume, one uses these single-site events uniformly distributed in the whole detector volume and obtained during WIMP search data taking. The fiducial volume is calculated from the ratio of the events in the triplet at 8.98, 9.66 and 10.37~keV surviving the fiducial cuts to the total of these events.
\begin{figure}
\begin{center}
\includegraphics[width=1\textwidth]{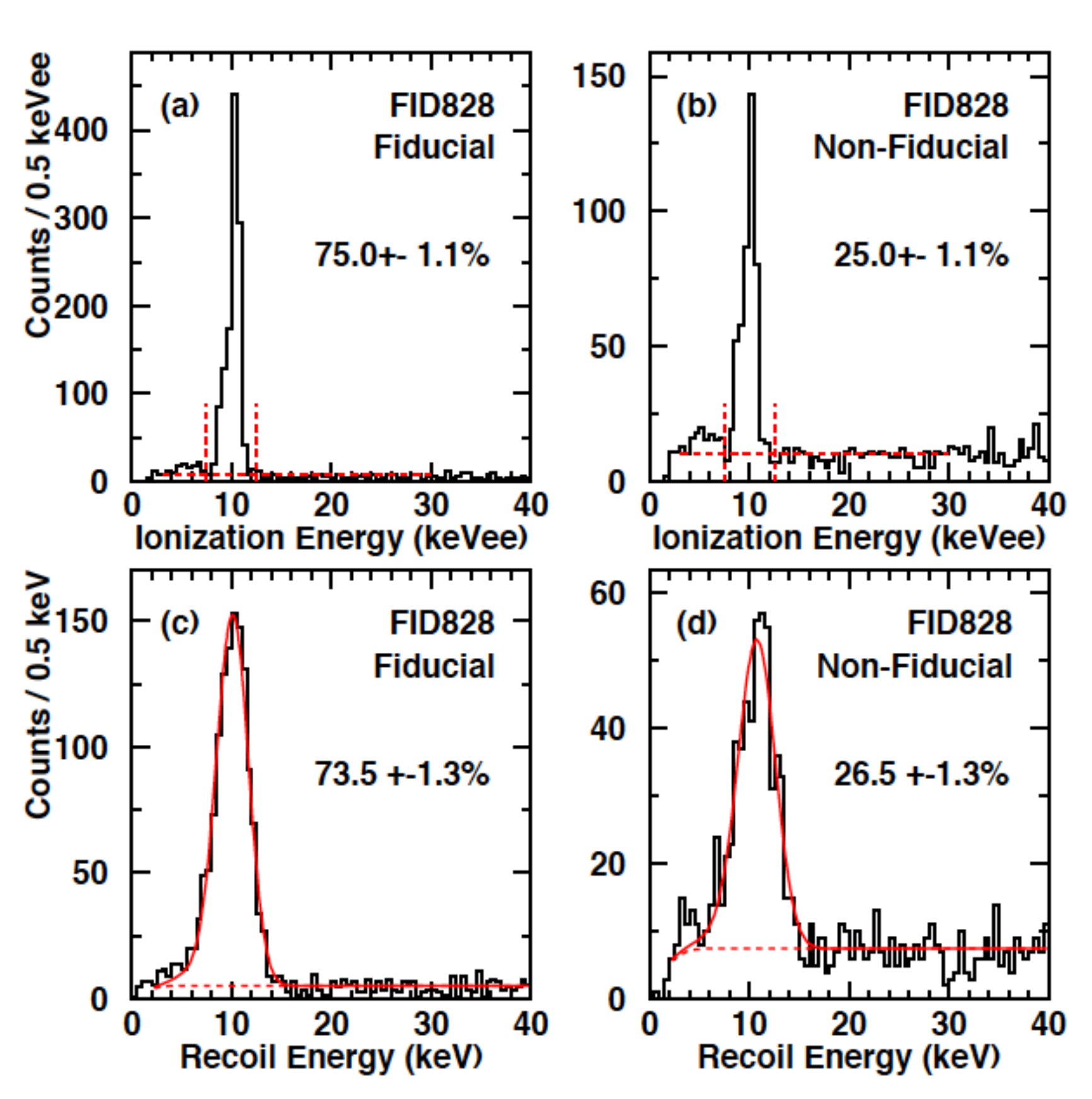}
\caption{Example of fiducial volume determination for the FID828 detector. Figures (a) and (c) (resp. (b) and (d)) show total ionization energy and recoil energy spectra for the fiducial events (resp. non-fiducial events), obtained after applying $0.55 <Q(E_r)<2.5$ and $E_i > 5 \sigma_{E_i}$ cuts. Associated fiducial volume (in \%) as obtained by the two methods described in the text are indicated in (a) and (c) plots.}
\label{fig:fiducialvolume}
\end{center}
\end{figure}

Cosmogenic peaks at 8.98 and 10.37 keV are clearly visible for most of the detectors, as seen in figures~\ref{fig:24-10keV-ion} and~\ref{fig:24-10keV-heat}, respectively, for fiducial ionization energy and combined heat energy.
Using the three peaks at 8.98~keV, 9.66~keV and 10.37~keV,
the first step of the fiducial volume determination consists of a fit of these lines to gaussian distributions in order to determine the relative fraction of events in each peak of the triplet.
A cut on the ionization yield value (see eq.~\ref{Q-2}), $0.55 < Q(E_r) < 2.5$, rejects $^{206}$Pb recoils and most of $\beta$ events, while keeping more than 96\% of $\gamma$-ray events in the first 2~mm under the electrodes. Furthermore, a cut on total ionization of $E_i > 5 \sigma_{E_i}$ around baseline is applied to reject heat-only events characterized by no ionization signal~\cite{Lukas}.

The 10~keV triplet can be observed in energy spectra both in ionization energy and recoil energy, for events accepted or rejected by fiducial cuts as shown in figure~\ref{fig:fiducialvolume}. To better control systematic effects due to background subtraction, two methods are used in order to evaluate the number of events in the peaks in the spectra with and without fiducial selection.  The first method consists of counting the number of fiducial events (non-fiducial events) in the energy interval $E_i$ between 7 and 13~keV$_{ee}$ after subtraction of the average background estimated between 3 and 30~keV$_{ee}$ (figure~\ref{fig:fiducialvolume} (a) and (b)). The second method consists of fitting the $E_r$ recoil energy spectrum with three gaussian functions with the same width and at relative positions and heights constrained by the values obtained in the fit of the fiducial data described above. The background is a constant multiplied by an error function (Erf) to take into account the efficiency loss due to the 5$\sigma$ cut on $E_i$ (figure~\ref{fig:fiducialvolume} (c) and (d)). In both methods, the fiducial volume fraction is the ratio of the number of fiducial events divided by the sum of fiducial and non-fiducial events. Results are presented in Table~\ref{tab:fiducialvolume} for 22 detectors.\footnote{FID822 (889~g) and FID840 (878~g) detectors are not included, since their fiducial volume measurements are not relevant (fiducial $B$ electrode of FID822 had a malfunctioning; $A$ and $C$ veto electrodes of FID840 had a readout problem leading to a planar configuration for this detector).}
The quoted values are the average obtained using both methods, the difference being used as a systematic uncertainty.

One can see from Table~\ref{tab:fiducialvolume} that 19 of the 22 bolometers have their fiducial volume fraction compatible within errors with a weighted average value of $(74.6 \pm 0.4)\%$, with a reduced $\chi^2$ of 1.04.  Three bolometers have lower fiducial volume fractions. Detector FID824 has a spacing between electrodes of 4~mm instead of 2~mm. The FID average thickness of the surface region for a spacing of 2 mm between electrodes (red region on figure~\ref{fig:FID800sheme_1}) is $\sim 2.5$~mm. It can be expected to double as the spacing doubles, nearly corresponding to the observed increase of non-fiducial volume in FID824. For the other two detectors (FID807 and FID826) the reduction of the volume fraction appears to be associated with wiring issues on the electrodes, which prevent them to be used for precision measurements. In these detectors some fiducial $\gamma$-ray events have ionization yield $Q(E_r) = 0.1$ probably due to a disconnected ring in a fiducial electrode. The analyses of Refs.~\cite{low-mass-paper, tritiumpaper, Lukas} use the fiducial volume fractions of Table~\ref{tab:fiducialvolume} in order to account for the value of FID824, and the possible $\pm 5\%$ detector-to-detector variation observed in that table.
\begin{table}[h]
\begin{center}
\caption{\label{tab:fiducialvolume}Fiducial volume fraction and associated total uncertainty, in \%, for 22 FID detectors. The quoted errors are the quadratic sum of systematic and statistic errors.}
\begin{tabular}{|c|c|c|}
\hline
Detector & Mass (g) & Fiducial fraction (\%) \\ \hline
FID807	& 820 & $55.6	\pm 5.9$ \\ \hline
FID810	& 820 & $73.2	\pm 3.4$ \\ \hline
FID817	& 878 & $ 78.3	\pm 1.6$ \\ \hline
FID820	& 877 & $72.2 \pm 3.3$ \\ \hline
FID821	& 878 & $75.0 \pm 3.4$ \\ \hline
FID823	& 882 & $72.7 \pm 5.6$ \\ \hline
FID824	& 877 & $62.5 \pm 2.3$\\ \hline
FID825	 & 878 & $74.3 \pm 2.3$ \\ \hline
FID826	 & 874 & $49.2\pm 7.2$ \\ \hline
FID827	 & 873 & $74.9 \pm 1.1$ \\ \hline
FID828	 & 871 & $74.4 \pm 1.8$ \\ \hline
FID831	& 878 &  $73.2	\pm 2.7$ \\ \hline
FID832	 & 875 & $70.4 \pm  5.2$ \\ \hline
FID837	& 875 & $72.7 \pm 1.4$ \\ \hline
FID838	&869 & $73.8 \pm 1.1$ \\ \hline
FID839	 & 869 & $76.1 \pm 1.1$ \\ \hline
FID841	& 878 & $74.3	\pm 1.3$ \\ \hline
FID842	 & 878 & $71.3	\pm 7.5$ \\ \hline
FID843	& 880 & $70.0	\pm 2.3$ \\ \hline
FID844	& 875 & $79.6	\pm 3.4$ \\ \hline
FID845	& 886 & $71.7	\pm 9.0$ \\ \hline
FID846	& 868 & $76.0	\pm 1.8$ \\ \hline 		
\end{tabular}
\end{center}
\end{table}

\subsection{$\gamma$-ray rejection factor}
\label{sec:gamma-rejection}
Using $\gamma$ calibrations of previous generation ID detectors performed in the context of EDELWEISS-II with a  $^{133}$Ba source, 6 events out of a total of $3.45\times 10^5$ were found with a reduced charge signal above a 20~keV threshold, which could mimic a nuclear recoil. A likely explanation for these events is the large region of low guard field of the ID detector design: misidentified $\gamma$-rays could originate from nonrejected multiple scatterings of $\gamma$ in the bulk (full charge collection) and in the planar electrode on the side of the detector (incomplete charge collection). The associated probability of $\gamma$-misidentification was calculated to be $R_{\gamma{\rm -mis-ID}} =(3 \pm 1) \times 10^{-5}$, leading to an expected background from $\gamma$-ray leakage of 0.9 events for the 384~$\rm{kg \cdot d}$ effective exposure of EDELWEISS-II~\cite{edw2}.

\begin{figure}[ht]
\begin{center}
\includegraphics[width=1.\textwidth]{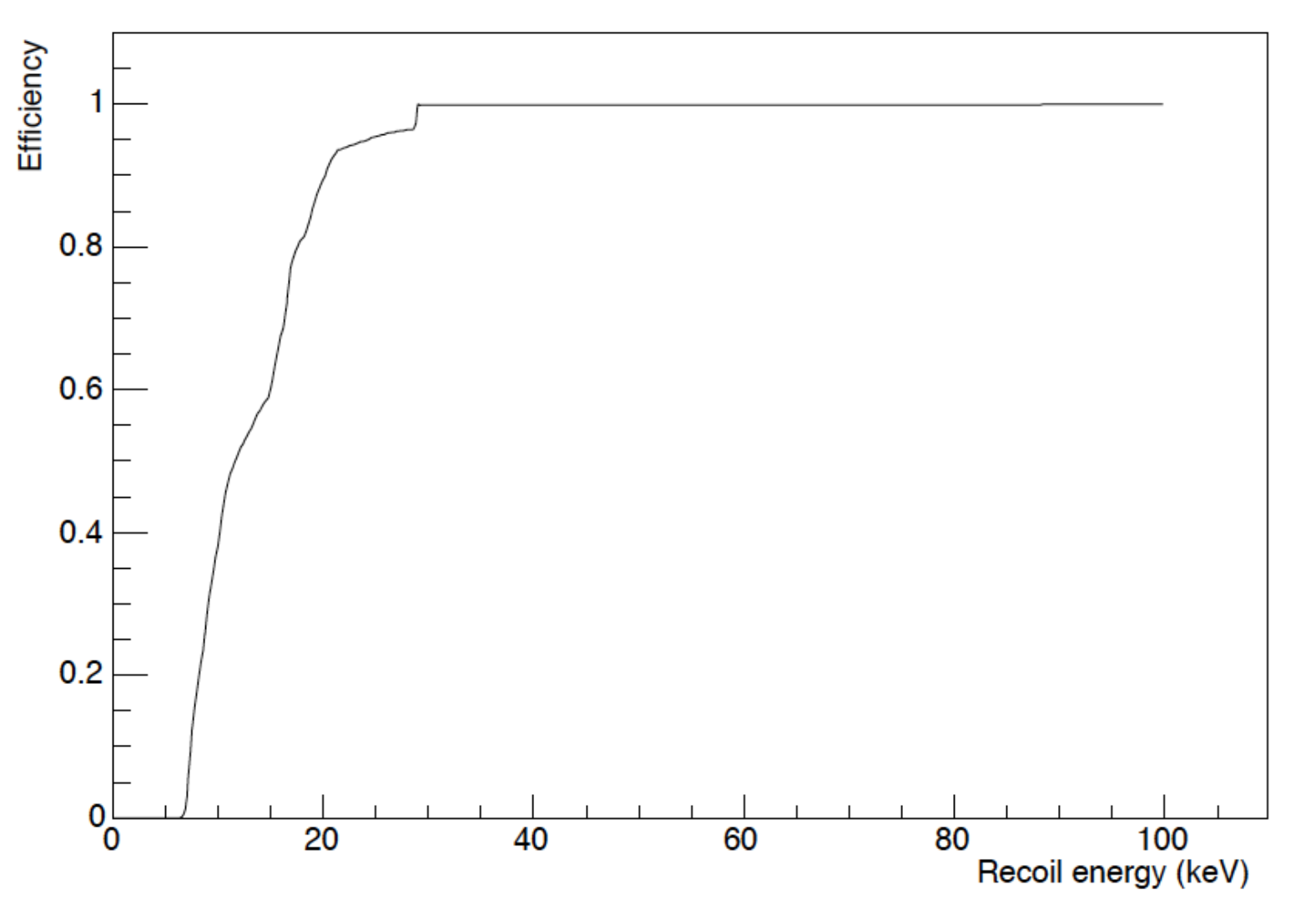}
\caption[]{Efficiency curve associated with data presented in figure~\ref{fig:gamma-rejection} and resulting from the combination of two different sets of data with FID detectors in the EDELWEISS setup.}
\label{fig:efficiency}
\end{center}
\end{figure}

\begin{figure}
\begin{center}
\includegraphics[width=1.\textwidth]{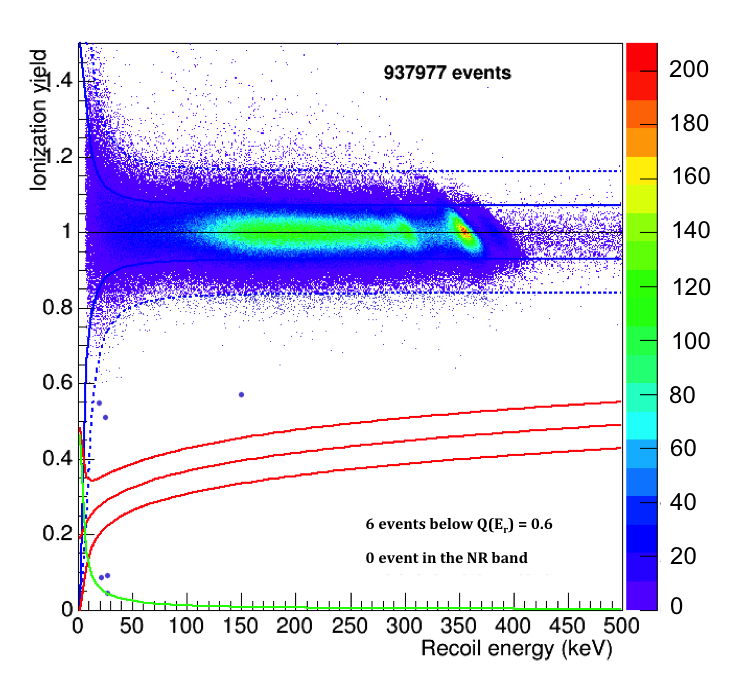}
\caption{Ionization yield $Q(E_r)$ versus recoil energy $E_r$ for events from $\gamma$ calibrations with $^{133}$Ba  sources with FID detectors.
Dashed and continuous blue lines correspond to 99.98\%\,C.L. and 90\%\,C.L. electronic recoil bands, whereas continuous red lines correspond to the 90\%\,C.L. nuclear recoil band. The green line corresponds to the fiducial ionization threshold. Although there are 6 single events below $Q(E_r) = 0.6$, none of the 937977 recorded events leaked into the nuclear recoil band.}
\label{fig:gamma-rejection}
\end{center}
\end{figure}

In the new FID detector design, where all planar electrodes are replaced by interleaved electrodes, this background of misidentified $\gamma$ is reduced by two effects: firstly the relative probability of Compton scattering in the non-fiducial volume is reduced by the reduction of this volume, and secondly, charge collection in this volume is improved as the electric fields in the non-fiducial region are systematically higher than in the bulk. Preliminary results on an improvement in $\gamma$ rejection had been reported in~\cite{AlexJ}. We present here an updated measurement, combining this result with additional $\gamma$ calibration data recorded with $^{133}$Ba sources, using the array of 24 FID detectors, more than doubling the total available statistics. The efficiency for detecting nuclear recoils in the combined analysis is shown in figure~\ref{fig:efficiency}, following the procedure described in~\cite{low-mass-paper}. It is 38\% at 10~keV and 89\% at 20~keV. The irregular rise of the function reflects the varying analysis thresholds of the different detectors and data acquisition periods.
Out of the $9.38 \times 10^5$ $\gamma$-ray events in the sample, shown in figure~\ref{fig:gamma-rejection}, six events are found below $Q(E_r)= 0.6$ but
none are observed in the 90\% C.L.\ nuclear recoil band. Information about the six events are reported in Table~\ref{tab:event-gamma}: the first three events were recorded during the 2010 calibration campaign by two of the first FID detectors, whereas the three last events are from Run 308 (see Table~\ref{tab:baseline-resol-threshold}).\footnote{The first three events have been recorded with FID803 and FID804 detectors. At the time of the measurement, their fiducial ionization resolutions were between 1.5 and 1.7~keV FWHM, and their heat resolutions were between 0.8 and 1.0~keV FWHM.} No systematic effects are seen, with events occurring in five different detectors and six different days.

\begin{table}[ht]
\caption{\label{tab:event-gamma}List of events with $Q(E_r) <0.6$ in figure~\ref{fig:gamma-rejection}. There is no obvious systematic effect since the six events are associated with different detectors or different dates.}
\begin{center}
\begin{tabular}{|c|c|c|c|c|}
\hline
Event & $E_r$  &$Q(E_r)$ & FID  & Event \\
number & (keV) & & Detector & date \\ \hline \hline
1 & 19.33 & 0.549 & 804 & Sep. 11$^{\rm th}$ 2010 \\ \hline
2 & 150.26 & 0.573 & 804 & Oct. 7$^{\rm th}$ 2010 \\ \hline
3 & 25.84 & 0.512 & 803 & Nov. 20$^{\rm th}$ 2010 \\ \hline
4 & 27.51 & 0.044 & 824 & Sep. 15$^{\rm th}$ 2014 \\ \hline
5 & 27.69 & 0.091 & 821 & Nov. 25$^{\rm th}$ 2014 \\ \hline
6 & 21.08 & 0.089 & 838 & Nov. 26$^{\rm th}$ 2014 \\ \hline 	
\end{tabular}
\end{center}
\end{table}
The 90\% C.L. limit on the leakage of $\gamma$-ray events above threshold into the nuclear recoil band is $R_{\gamma{\rm -mis-FID}} < 2.5 \times 10^{-6}$, at least a factor 12 better than the leaking fraction measured with ID detectors.

\subsection{Surface rejection factor}
\label{sec:surfacerejection}
\begin{figure}[ht]
\begin{center}
\includegraphics[width=1.\textwidth]{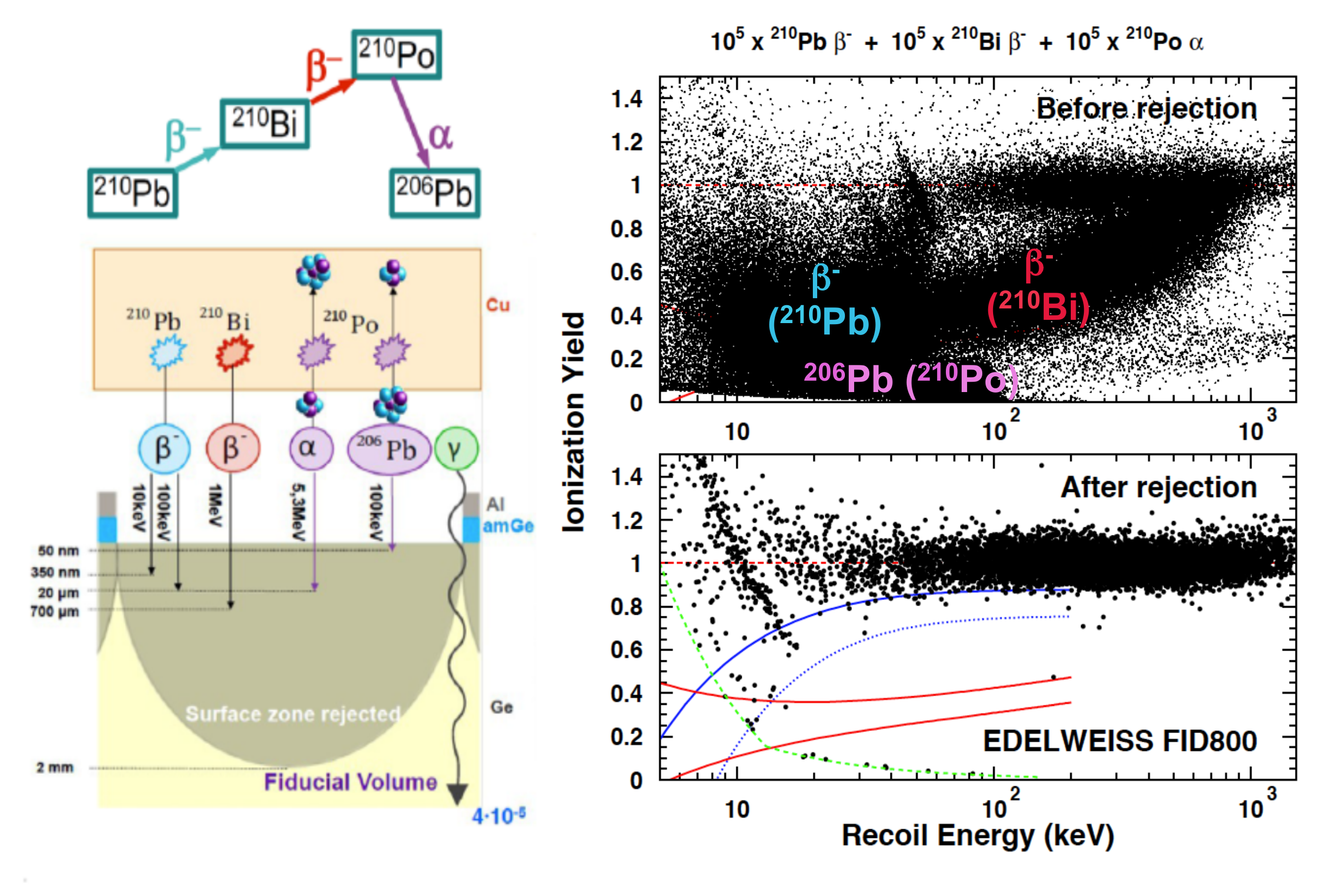}
\caption[]{Left: schematic of the surface calibration with a $^{210}$Pb source which decays, emitting $\beta$- and $\alpha$-particles of various energies. Right: distribution of ionization yield as a function of recoil energy observed in two 800-g FID detectors, for an exposure to $10^5$ $^{210}$Pb decays from a source where this isotope is close to equilibrium with its $^{210}$Bi and $^{210}$Po
daughters before (top) and after (bottom) fiducial cuts. The band where 90\% of all nuclear recoil events are expected at an ionization yield close to 0.3 is shown in red. The full and dashed blue lines indicate the 90\% and 99.98\%\,C.L.\ regions for $\gamma$-rays, respectively. The green dashed hyperbola shows the 2~keV ionization threshold cut. The various populations are indicated accordingly to the color code used in the left schematic. Only one event at 15~keV is remaining inside the nuclear recoil band after fiducial cut~\cite{Gascon-Bastidon}.}
\label{fig:surface-rejection}
\end{center}
\end{figure}

Events occurring at the detector surface suffer from incomplete charge collection, which leads to an underestimation of the ionization yield $Q(E_r)$. The main surface background source is radon daughters deposited on the detector surface and the copper housing. Radon ($^{222}$Rn, $T_{1/2} = 3.84$~d) quickly decays (via a series of decays) into the long-lived isotope of $^{210}$Pb ($T_{1/2} = 22.24$~years). As shown in figure~\ref{fig:surface-rejection} (left), the decay of $^{210}$Pb leads to the stable $^{206}$Pb isotope via the cascade:
\begin{equation}
 ^{210}{\rm Pb} \xrightarrow[Q~=~63.5~{\rm keV}]{\beta^-} ~^{210}{\rm Bi} \xrightarrow[Q~=~1162.7~{\rm keV}]{\beta^-} ~^{210}{\rm Po} \xrightarrow[Q~=~5407.5~{\rm keV}]{\alpha} ~^{206}{\rm Pb}
\label{Pb-decay-chain}
\end{equation}
This cascade results in the emission of both low- and high-energy $\beta$-particles, an $\alpha$-particle of 5.3~MeV and a 103~keV recoiling $^{206}$Pb nucleus. These emissions are nearly in equilibrium if the contaminated material is significantly older than the $T_{1/2} = 138$~d of $^{210}{\rm Po}$. The surface rejection factor of ID detectors was measured in the context of EDELWEISS-II using dedicated $^{210}$Pb calibrations and was found to be $R_{\rm surf-ID} =6 \times 10^{-5}$ events per $\alpha$ at 90\%\,C.L. for a 20~keV threshold~\cite{edw2}.

In order to study the 800-g FID bolometer response to $\alpha$- and $\beta$-back\-grounds, a dedicated calibration has been carried out in 2012 with two of the first detectors.\footnote{The detectors are FID803 and FID808. At the time of the measurement, their fiducial ionization resolutions were between 0.7 and 1~keV FWHM, and their heat resolutions were between 1 and 1.4~keV FWHM.} For a specific data taking period, they have been equipped with a $^{210}$Pb source, fabricated by exposing copper adhesive tape to a radon source. This tape was then attached to the inner surface of the copper case of the Ge detector in such a way that it faced all detector surfaces. As previously, the rejection power of the FID detectors against surface events coming from $^{206}$Pb recoils, $\beta$ or $\alpha$ backgrounds is measured from the number of events that are observed before and after surface rejection.

Depending on the particle type and energy, the particles associated with the $^{210}$Pb decay penetrate the detector volume to various depths (see figure~\ref{fig:surface-rejection}, left), with either complete or incomplete charge collection. Most of these events should be rejected by applying fiducial cuts (section~\ref{sec:fiducialcuts}). By counting the number of  remaining events after applying these cuts, the number of misidentified surface events is determined: one event has been found at 15~keV in the 90\%\,C.L. nuclear recoil band for an exposure of $9.7 \times 10^4$ $\alpha$-particles, as shown in figure~\ref{fig:surface-rejection} (right), where the remaining events with an ionization yield of $\sim 1$ are mostly due to the $\gamma$-ray background in the experimental site and to cosmogenic
activation of the germanium crystal (section~\ref{sec:sub:fiducialvolume}).
From this observation, a 90\%\,C.L. limit on the surface event leaking factor $R_{\rm surf-FID} < 4 \times 10^{-5}$ is derived for a recoil energy threshold of 15~keV~\cite{Gascon-Bastidon}. This limit is a factor 1.5 better with respect to the ID design and is measured at lower threshold.

The 46~keV $\gamma$-rays emitted in the $\beta$-decay of $^{210}{\rm Pb}$ with a branching ratio of 4.25\% and a penetration depth of 0.4~mm in germanium appear with $Q(E_r) = 1$ on figure~\ref{fig:surface-rejection} (top right). These $\gamma$-rays are essentially removed by the fiducial cut (figure~\ref{fig:surface-rejection} (bottom right)), consistent with the expectation that most of them are contained within a depth of 2~mm under the surfaces (section~\ref{sec:sub:fiducialvolume}).

The cleaning procedure, as described in section~\ref{sec:detectors}, allows to reduce surface $\alpha$ contamination due to radon daughters. However it is possible to measure the total rate of $\alpha$-particles per day for each of the 24 detectors used for coincidence studies (FID detector masses are given in  Table~\ref{tab:fiducialvolume}). They appear as a peak of events with $E_r = 5.3$~MeV accompanied by a tail of $\alpha$-particles with degraded energies at $Q(E_r) \sim 0.25$ that can be observed on the top right panel of figure~\ref{fig:surface-rejection}. The rates vary from $2.00 \pm 0.11$ to $10.15 \pm 0.26$ $\alpha$/d. With the detector mass as given in~\ref{tab:fiducialvolume} this leads to an average total rate of 5.45~$\alpha$/kg/d, with a standard deviation of $\sigma=2.9$. Thus less than one surface event per fiducial exposure of 3\,400~$\rm{kg \cdot d}$  is expected in the NR band, considering a fiducial volume fraction of 74.6\% (section~\ref{sec:sub:fiducialvolume}).

\section{Performance of the EDELWEISS-III detector array}
\label{sec:array-performance}

Despite active and passive shieldings, there remains a small residual neutron background which can mimic WIMPs in the nuclear recoil band. It originates from high energy neutrons produced by muons that are not tagged by the veto system and from radiogenic neutrons from $^{238}$U and $^{232}$Th $(\alpha, n)$ reactions and spontaneous fission in the materials surrounding the detectors.

The contamination activity measurements and upper limits associated to the radiopurity budget of the experiment are reported in Table~\ref{table-radiopurity-measurements}. The quoted activities of $^{238}$U, $^{232}$Th and $^{210}$Pb assume that all daughters in their chains are in equilibrium. Activities for the $^{232}$Th chain are measured using $\gamma$-rays from $^{228}$Ac, $^{212}$Pb,  $^{212}$Bi or $^{208}$Tl, whereas for the $^{238}$U chain, $\gamma$-rays from $^{214}$Pb, $^{214}$Bi or $^{234m}$Pa are used. The large difference between activity values obtained for $^{238}$U and $^{210}$Pb can be due to contamination of surfaces with $^{210}$Pb.

A special attention was devoted to the activities of the Mill-Max connectors which are the closest to the germanium detectors.  Each connector is made of three different components: a pin and a socket, in brass, associated with a copper-beryllium press-fit contact. The activities of these components are listed in the first three lines of Table~\ref{table-radiopurity-measurements}.  The two brass pieces revealed a specific $^{210}$Pb contamination corresponding to a total activity of $\sim 0.3$~Bq. Also the press-fit contact, even representing less than 10\% of the Mill-Max connector mass, is the most important source of $^{238}$U close to the detectors, with a total activity of $\sim 20$~mBq. The precise measurement of U/Th activities for the three connector pieces were performed in 2015 using ICP-MS. At that time there was no available alternative to replace these connectors while still providing a highly reliable contact at 18~mK.

\begin{sidewaystable}[htbp]
\caption{\label{table-radiopurity-measurements}The source of background radiation and its mass in the EDELWEISS-III setup as used in the simulations. Decay rates, in mBq/kg, for each radioactive isotope considered are listed in the other columns. The quoted activities of $^{238}$U and $^{232}$Th assume that all daughters in their chains are in equilibrium. All upper limits are given at 90\%\,C.L.. If not mentioned, the measurements have been performed using Gentiane HPGe detector. Other measurements are: $^{(1)}$~U/Th activities measured with ICP-MS in 2015 at ENS-Lyon; $^{(2)}$~(GeMPI4) and $^{(3)}$~(GeMPI2) by M. Laubenstein (2014)~\cite{Laubenstein}.  Abbreviations are used when needed: "conn." for connectors, "cont." for contacts. Also, $^{\rm (a)}$ CuC2 copper from Carlier~\cite{carlier}; $^{\rm (b)}$ cables for connecting 100~K and 1~K volumes; $^{\rm (c)}$ the cryostat structure is made of stainless steel; $^{\rm (d)}$ cables for connecting 300~K and 100~K volumes.}
\begin{center}
\begin{tabular}{|l|c|c c |c |c c c|}
\hline \hline
\multicolumn{1}{|c|}{} &\multicolumn{1}{|c|}{Mass} & \multicolumn{2}{|c|}{$^{238}$U chain} & \multicolumn{1}{|c|}{$^{232}$Th chain} & \multicolumn{1}{|c}{} &\multicolumn{1}{c}{} &\multicolumn{1}{c|}{}\\
\cline{3-5}
\multicolumn{1}{|c|}{Element} & \multicolumn{1}{|c|}{ (kg)} & \multicolumn{1}{|c}{$^{210}$Pb} & \multicolumn{1}{c|}{$^{238}$U} &  \multicolumn{1}{|c|}{$^{232}$Th} & \multicolumn{1}{|c}{$^{40}$K} &\multicolumn{1}{c}{$^{60}$Co} &\multicolumn{1}{c|}{$^{137}$Cs} \\
\hline \hline
Mill-Max conn.$^{(1)}$ & & & & & & & \\
Brass pins & 7.39 g & $(1.1\pm 0.1)\times 10^4$ &$<62$ & $<20$  & $675	\pm 221$ & $<36$ & $< 47$ \\
\cline{2-8}
Brass sockets & 9.41 g& $(2.6	\pm 0.4)\times 10^4$ & $<62$ & $<20$ & $<2645$ & $<129$ & $< 132$ \\
\cline{2-8}
CuBe press-fit cont. & 1.63 g& & $(1.2	\pm 0.2)\times10^{4}$ & $980	\pm 196$ & &  &  \\
\hline \hline
PTFE Delrin  & 0.040& $< 26$ & $< 16$ & $1.5	\pm 1$ & $< 43$ & $< 2.3$ & $< 2.0$\\ \hline
Kapton connectors & 0.094 & $< 187$ & $14	\pm 7$ &  $67	\pm 31$& $150	\pm 98$& & \\ \hline
PTFE contacts & 0.061& & $10	\pm 5$ &  $20	\pm 7$& & & \\ \hline
Brass FID casings & 0.40& $524	\pm 102$ & $< 16$ & $< 15$ & $< 75$& $5\pm 3$ & $3\pm 2$\\ \hline
Cu Kapton cables & 0.51 & $549	\pm 111$ & $8\pm 6$ & $15\pm 10$ & $66\pm 26$& $3\pm 2$ & $< 4.0$\\ \hline
Brass screws & 2.0& $620\pm 254$ & & $3.5\pm 0.9$& $< 19$& $< 3.0$& $2.6\pm 1.5$\\ \hline
Cu (NOSV)$^{(2)}$ & 295& & $<0.39$ & $<32 .10^{-3} $ & $<0.15$ & $(35 \pm 9)\times 10^{-3} $ & $<6.4\times 10^{-3} $ \\ \hline
Cu (CuC2)$^{(3)}\,^{\rm (a)}$& 328& & $<40\times 10^{-3} $ & $(24	\pm 12)\times10^{-3} $& $<0.50$& $(42 \pm 16)\times10^{-3} $& $<35\times10^{-3} $\\ \hline
PE internal &151 & $< 3.0$& $0.65	\pm 0.08$& $0.30	\pm 0.07$& $< 1.0$& $< 0.06$& $< 0.06$\\ \hline
Axon cables$^{\rm (b)}$ & 3.5 & $138	\pm 53$& $4	\pm 3$& $5	\pm 2$& $177	\pm 22$& $< 5.0$& $< 2.0$\\ \hline
Other connectors$^{\rm (b)}$ & 0.43 & $(6.0\pm 0.5)\times10^{3} $& $(2.6\pm 0.4)\times10^{3} $& $450	\pm 44$& $<571$& $<36$& $<39$\\ \hline
Cryostat stucture$^{\rm (c)}$ & 1236& & $<1.0$ & $<1.0$& & & \\ \hline
PCB FET boxes & 0.55& $(1.4\pm 0.3)\times10^{4} $&  $(7.5\pm 0.2)\times10^{3} $& $(10.1\pm 0.1)\times10^{3} $& $(11.5\pm 0.6)\times10^3 $& & \\ \hline
PCB FID boxes & 10.4& & $<1660$& $<1215$& & & \\ \hline
Al FID boxes & 27.8& $88\pm 36$& $4\pm 3$& $<2.0$&$65\pm 34$ & $5\pm 3$& $2.0\pm 1.3$\\ \hline
Axon cables$^{\rm (d)}$ & 6.32& & $182\pm 70$&  $13.0	\pm 2.5$& & &  \\ \hline \hline
Mild steel structure & 8\,600 & &$<1.0$ & $<1.0$ & &   &  \\
\hline
Polyethylene & 40\,000 & &$<12$& $<0.4$ & $16	\pm 2$&  $<3.0$ &  \\
\hline
Lead & 39\,000 &  & $<1.0$ & $<1.0$ & &   &  \\
\hline
Rock~\cite{chazal} &  & &$(10\pm 3)\times10^{3} $ & $(10.0\pm 0.8)\times10^{3} $ & $(230\pm 30)\times10^{3} $ &   &  \\
\hline
Concrete~\cite{chazal} &  & &$(23\pm 3)\times10^{3} $ & $(5.7\pm 0.8)\times10^{3} $ & $(77\pm 13)\times10^{3} $ &   &  \\
\hline \hline	
\end{tabular}
\end{center}
\end{sidewaystable}

\subsection{Shielding performance against $\mu$-induced neutron backgrounds}
\label{sec:sub:peformancemuons}

Muon-induced events in germanium bolometers are rejected using the precise synchronisation with the $\mu$-veto system. In a dedicated analysis studying coincidences between the $\mu$-veto system and the bolometers, it was shown that the $\mu$-induced bolometer events can be distinguished from other backgrounds on the basis of their bolometer multiplicity and their total energy deposit in the bolometer. Selecting these events using the hits in the Ge bolometers only, a lower limit of the $\mu$-veto tagging efficiency of $\varepsilon_{\mu{\rm -veto}} > 93\%$ (90\%\,C.L.) was derived being limited by the scarce statistics. Geant4-based simulations of muons entering the modified geometry of the experiment were performed in order to derive the rate of all events in the bolometer array induced by muons before applying a veto cut. This rate has been shown to be in good agreement with the measured one extracted from the data~\cite{theseCecile}.

Selecting single fiducial events in the 90\%~C.L. nuclear recoil band with $E_r =$~[10--100]~keV, no coincidence with the $\mu$-veto was identified, leading to an upper limit of the measured rate of $\mu$-induced WIMP-like events in EDELWEISS-III of $\Gamma^{\rm WIMP-like}_{\mu{\rm -ind,meas}} < (1.7 \times 10^{-2})$ events/d (90\%~\,C.L.). This rate was found to be $\Gamma^{\rm WIMP-like}_{\mu{\rm -ind,simu}} = (0.76 \pm 0.01 (stat) ^{+0.26}_{-0.17} (syst))  \times 10^{-2}$~events/d in the corresponding simulation, being compatible with the above upper limit.

For an accumulated exposure of 600{$\, \rm{kg \cdot d}$} in the actual geometry of the bolometer array, the pre-veto-cut rate leads to an expected $\mu$-induced background of $N^{\rm WIMP-like}_{\mu{\rm -ind,simu}} =0.36 \pm 0.02 (stat) ^{+0.12}_{-0.07} (syst)$ events in the WIMP mass range [10--1000]~GeV. After applying the $\mu$-veto cut, an upper limit of $N^{\rm WIMP-like}_{\mu{\rm -ind}} < 0.06$ events (90\%\,C.L.) was derived.
These results demonstrate that $\mu$-induced background is negligible for the WIMP search analyses performed with the data of the EDELWEISS-III experiment. More details on the $\mu$-veto, on an equivalent analysis for a low-mass WIMP search in the range [3--30]~GeV, and on the measured $\mu$ flux can be found in~\cite{muonveto, theseCecile}.

\subsection{Shielding performance against radiogenic neutrons and $\gamma$-rays}

Preliminary $\gamma$ and radiogenic neutron background studies have been carried out~\cite{SilviaLRT} in order to show that the performance of the EDELWEISS-III environment and shieldings are adequate for dark matter searches.

Radiogenic neutron background, as described in section~\ref{subsec:radiogenic-neutrons}, has been studied by Monte Carlo (MC) simulations where several sources of neutrons were considered. In a first step neutron production rates in the materials of the internal parts of the detector and spectra from $(\alpha,n)$ reactions, spontaneous fission and delayed emission due to the decay of radionuclei have been determined using the SOURCES4A code~\cite{sources4a} which has been modified to extend the energies of $\alpha$-particles up to 10 MeV, and to improve and extend the ($\alpha$,n) cross-section library~\cite{tomasello}. Then Geant4.9.6~\cite{Geant4} has been used to propagate neutrons through the experimental setup, assuming secular equilibrium for uranium and thorium decay chains. The number of nuclear recoils observed in an extended region of interest for WIMP search ([10--200]~keV) in the fiducial volume and considering one year of data acquisition has been derived. Using an array of twenty-four FID detectors with a mean fiducial mass of 620~g each (see Table~\ref{tab:fiducialvolume}), the number of single (multiple) nuclear recoils in the 90\% NR band has been found to be 1.4 (3.4)  for an exposure of 5\,435~$\rm{kg \cdot d}$. The simulations in the restricted [10--100]~keV range also show a ratio of single over multiple scattering neutron events of 0.45, with a systematic uncertainty of 30\% depending on the origin of the neutrons. The spectral shape of the neutron-induced nuclear recoil distribution is relatively independent of the nature and location of the neutron background source.

\begin{figure}
\begin{center}
\includegraphics[width=1\textwidth]{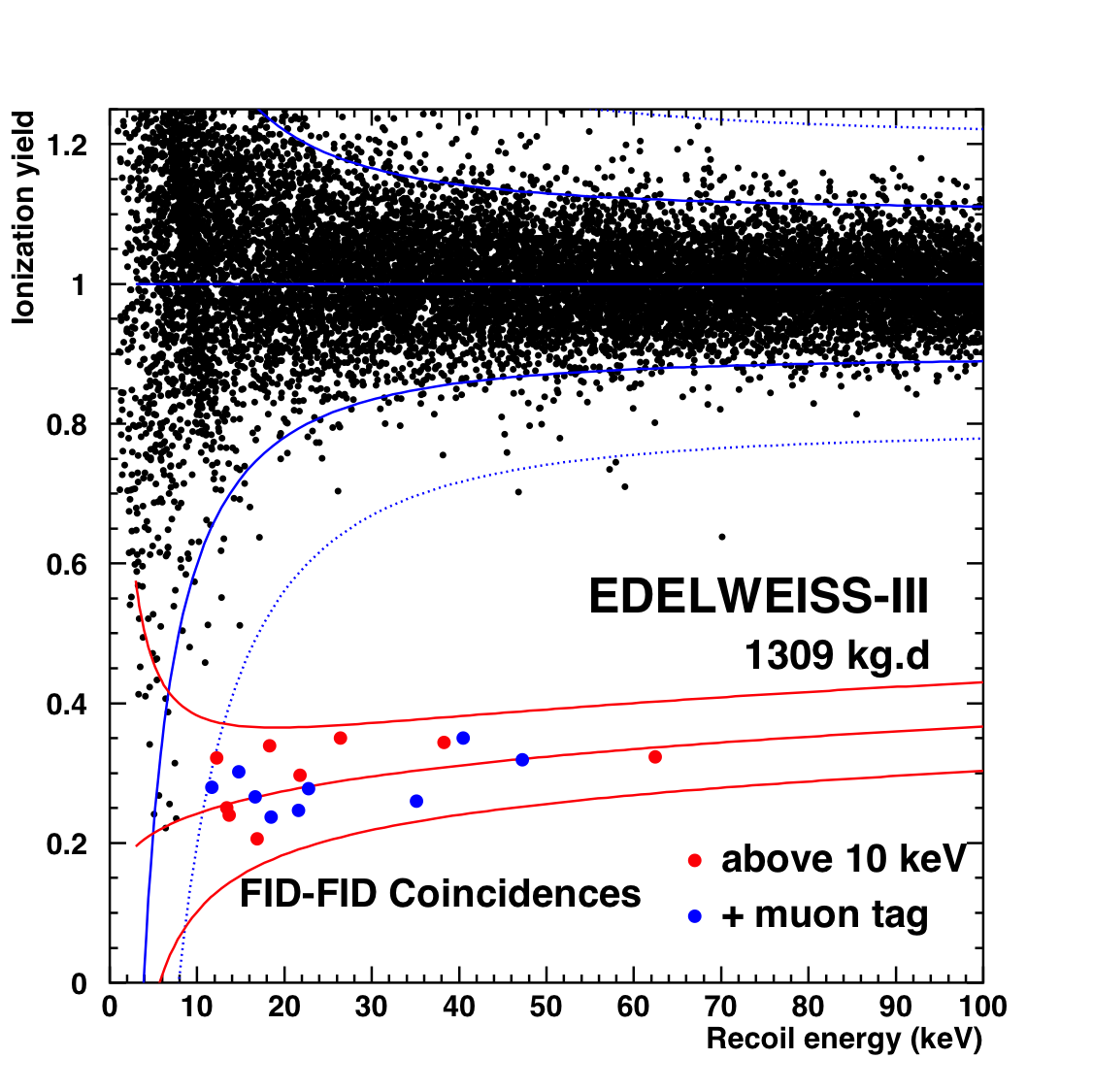}
\caption{Ionization yield versus recoil energy in the fiducial volume of individual detectors for the 1\,309~$\rm{kg \cdot d}$ effective exposure of Run 308, for hits in coincidence with a signal in at least one other FID detector. The multiplicity associated to each fiducial recoil is evaluated using the entire volume of all detectors. There are 18 such multiple hits in the nuclear recoil band in the [10--200]~keV range (colored circles), where nine of them are also in coincidence with the muon veto (in blue) and the nine others are not (in red).
The middle red curve shows the mean ionisation yield for the NR band whereas the red curves on both sides show the area where 90\% C.L. of nuclear recoil events are expected to lie. Similarly, the middle blue curve shows the average ionisation yield for electron recoils (taken as 1 by definition) whereas the two solid (dotted) curves on both sides show the 90\% (99.98\%) C.L.\ areas for electron recoil events.}
\label{fig:coinc-run308}
\end{center}
\end{figure}

More importantly the real neutron flux can be checked inside the EDEL\-WEISS-III detector array, independently from the actual WIMP search data, by using events where more than one detector have triggered simultaneously. Figure~\ref{fig:coinc-run308} shows multiple hit events in the NR region registered during the WIMP search data taking periods of Run 308, using all 24 available FID detectors (see Table~\ref{tab:baseline-resol-threshold}). Out of the 18 nuclear recoils in coincidence with another detector observed in the [10--200]~keV range for an exposure of 1\,309~$\rm{kg \cdot d}$, nine have been found in coincidence with the muon veto. The other nine recoils cannot be due to $\mu$-induced neutrons as demonstrated in section~\ref{sec:sub:peformancemuons}. They are multiple hits with ionization yields consistent with nuclear recoils in all detectors. This number exceeds the expectation for this exposure of 0.8 nuclear recoils associated to multiple hits deduced from the Monte Carlo simulation for this exposure. This excess could be due to defects in the PE shielding, currently under investigation. The observed number of multiple hits has been used to re-scale the radiogenic neutron flux used in the Monte Carlo predictions, as described in~\cite{low-mass-paper}.

Despite the lead shielding described in section~\ref{sec:shields}, most of the event rate in the FID detectors is due to $\gamma$-ray interactions.  An analysis of data from EDELWEISS-III in the energy range from 100~keV to 4~MeV (outside the ROI for dark matter searches) has been carried out leading to an average integrated $\gamma$ counting rate in individual detectors of $370 \pm 4$~$\gamma$/kg/d based on the exposure of about 554~$\rm{kg \cdot d}$ for the total detection volume. In the fiducial volume, the $\gamma$ rate in individual detectors in the same energy range is $235\pm 5$~$\gamma$/kg/d, corresponding to a fiducial exposure of about 380~$\rm{kg \cdot d}$. MC simulations with Geant4.9.6~\cite{Geant4} have been carried out to understand the experimental $\gamma$-background spectrum. All materials surrounding the detectors in the 10~mK and 1~K volumes along with the shielding have been considered as sources of $\gamma$-background with contaminations from uranium and thorium chains, as well as $^{40}$K, $^{60}$Co and $^{137}$Cs isotopes. They have been identified either by $\gamma$-spectroscopy in dedicated HPGe screening measurements or by ICP-MS analyses (see section~\ref{sec:shields} and Table~\ref{table-radiopurity-measurements} in this section).

These contaminations are taken as inputs for simulations. A preliminary comparison of data and simulations has been carried out and a relatively good agreement is observed for the ionization spectrum in both the total volume and the fiducial volume, as described in~\cite{SilviaLRT}.  The measured $\gamma$-background in the fiducial volume is 70~$\gamma$/kg/d in the range [20--200]~keV, compared to a predicted value of 78~$\gamma$/kg/d. The dominant source of this background are the Mill-Max connectors with their press-fit contacts in CuBe glued on the FID detector casings, as already explained at the beginning of section~\ref{sec:array-performance}.\footnote{See information on the Mill-Max connectors in
https://www.mill-max.com/engineering\_notebooks/detail/83.}
More detailed investigations on neutron and $\gamma$ backgrounds, both with latest versions of neutron and $\gamma$ MC simulations, are ongoing~\cite{new-gamma-bkg-publi}.

\section{Summary and prospects}
\label{sec:conclusion}

In this paper, we reported the design and performance of the EDELWEISS-III dark matter direct detection search experiment operating in the Modane Underground Laboratory using an array of 24 FID cryogenic germanium bolometers. This third phase of the experiment involves upgrades on both the electronics (readout and DAQ) and the cryogenic systems, with new shieldings having been installed. These measures, coupled with the use of the newly designed FID detectors, allowed to improve energy resolutions and thus lower the thresholds. The characterization of the detectors and their environment was performed on data sets separated from blinded data for the WIMP search analysis, notably multiple hit events, events outside the fiducial volume and fiducial events with an ionization yield larger than 0.5 with respect to the mean ionization yield from electron recoils. It was based mainly on data obtained during the 10-months Run 308, but also on data from shorter data-taking periods and special calibration data.

During Run 308, the fiducial ionization energy resolutions were typically below 0.7 keV$_{ee}$ FHWM whereas the resolution values were typically below 1.5 keV$_{ee}$ FWHM for heat energies for the whole set of 24 FID detectors used for coincidence studies. A subset of eight FID bolometers with clearly better baseline resolutions less than 0.7~keV$_{ee}$ and 1~keV$_{ee}$ for the fiducial ionization FWHM$_{fid}$ and the combined heat baseline FWHM$_{c}$ respectively, has been identified. A low-enough online threshold of less than 1.5~keV$_{ee}$ has been achieved to perform low WIMP mass studies. Also the fiducial volume fraction of the FID detectors has been measured using the activation of the $^{65}$Zn and $^{68,71}$Ge isotopes, with a weighted average value of $(74.6 \pm 0.4)\%$.

The capability of the FID germanium detectors of clearly separating the electron recoils induced by $\beta$ and $\gamma$ radiation from nuclear recoils using a double-readout event-by-event discrimination has been quantified. The rejection of bulk $\gamma$-ray events has been measured with $\gamma$-calibration data and the limit has been found to be $R_{\gamma{\rm -mis-FID}} < 2.5 \times 10^{-6}$ at 90\%\,C.L., at least a factor 12 better than the leaking fraction measured with the previous generation of detectors. Also the FID detectors allow an efficient rejection of surface $\beta$- and $\alpha$-decays. A value of $R_{\rm surf-FID} < 4 \times 10^{-5}$ (90\%\,C.L.) for $\alpha$-events has been measured using data from a $^{210}$Pb source, which demonstrates that less than one surface event per fiducial exposure of 3\,400~$\rm{kg \cdot d}$ is expected leaking in the NR band for a recoil energy threshold of 15~keV.

The efficiency of the $\mu$-veto, which acts as an active shielding against cosmogenic $\mu$-induced neutrons, has been studied. An upper limit of $N^{\rm WIMP-like}_{\mu{\rm -ind}}$ $< 0.06$ events (90\%\,C.L.) was derived for an exposure of 600{$\, \rm{kg \cdot d}$}. It demonstrated that the number of $\mu$-induced background events in the nuclear recoil band is negligible for the EDELWEISS-III WIMP search analyses. Performance of the passive shieldings against neutrons and $\gamma$-rays were also discussed together with the current levels of natural radioactivity measured in the experiment. The measured $\gamma$-background in the fiducial volume is 70~$\gamma$/kg/d in the range [20--200]~keV.  An unexpected excess of neutron-induced nuclear recoil events (multiple hits) has been observed and is currently under investigation.

These results demonstrate that all elements of the EDELWEISS-III setup have been tested and have exhibited a performance sufficient for low-mass WIMP searches. Extension of the search to higher WIMP mass is only limited by the neutron background inside the cryostat, as the achieved resolutions, rejection capabilities and thresholds of the upgraded EDELWEISS-III detectors are excellent.

Having presented first results on WIMP search~\cite{low-mass-paper, Lukas}, EDELWEISS is now focusing on the optimization of the experiment to enhance its sensitivity for low-mass WIMPs through four R\&D tasks~\cite{papier-projections}. The first one is to benefit from the so-called Neganov-Luke boosting effect to lower the energy thresholds. For this purpose upgrades on readout electronics to allow a high ionization biasing up to $\pm$70~V have already been made. Detector surface treatments have been done to ensure small leakage currents at high voltages. Meanwhile, R\&D is being carried out on detectors with three other tasks: two of them in order to improve baseline energy resolutions, both in heat and fiducial ionization, and the third one in order to understand the origin of heat-only events to reduce their rate significantly.

\section{Acknowledgments}


The help of the technical staff of the Laboratoire Souterrain de Modane and the participant laboratories is gratefully acknowledged. The EDELWEISS project is supported in part by the German ministry of science and education (BMBF Verbundforschung ATP Proj.-Nr.~05A14VKA), by the Helmholtz Alliance for Astroparticle Physics (HAP), by the French Agence Nationale pour la Recherche (ANR) and the LabEx Lyon Institute of Origins (ANR-10-LABX-0066) of the Universit\'e de Lyon within the program ``Investissements d'Avenir'' (ANR-11-IDEX-00007), by the P2IO LabEx (ANR-10-LABX-0038) in the framework ``Investissements d'Avenir'' (ANR-11-IDEX-0003-01) managed by the ANR (France), by Science and Technology Facilities Council (UK), and the Russian Foundation for Basic Research (grant No. 07-02-00355-a).

\newpage


\end{document}